\documentclass[lettersize,journal]{IEEEtran}
\usepackage{amsmath,amsfonts,amssymb}
\usepackage{algorithmic}
\usepackage{algorithm}
\usepackage{array}
\usepackage{eurosym}
\usepackage{diagbox}

\usepackage{caption}
\usepackage{float}
\usepackage{textcomp}
\usepackage{stfloats}
\usepackage{url}
\usepackage{enumerate}
\usepackage{nccmath} 
\usepackage{verbatim}
\usepackage{graphicx}
\usepackage{cite}
\usepackage{graphicx}
\usepackage{textcomp}
\usepackage{xcolor}
\usepackage{bm}
\usepackage{subcaption}
\makeatletter \renewcommand{\maketag@@@}[1]{\hbox{\m@th\normalsize\normalfont#1}}
\usepackage{subfloat}
\usepackage{balance}
\usepackage{autobreak}

\usepackage{url}
\usepackage{caption}
\usepackage{cite}
\usepackage {epstopdf}
\usepackage{orcidlink} %
\hypersetup{hidelinks} 

\newcommand{\avector}[2]{\boldsymbol{a}_{#1}^{#2}(\varphi_{#1}^e,\varphi_{#1}^a,\tilde{\boldsymbol{#2}})}

\begin{document}
	
	\title{\huge Robust Transceiver Design for RIS Enhanced Dual-Functional Radar-Communication with Movable Antenna}
	
	\author{Ran Yang, \textit{Graduate Student Member, IEEE,} Zheng Dong, \textit{Member, IEEE,} Yue Xiu, \textit{Member, IEEE},\\ Guangyi Liu, \textit{Senior Member, IEEE}, Wanting Lyu, \textit{Graduate Student Member, IEEE,}  \\  Xiangxin Meng, Yan Li, \textit{Member, IEEE}, and Ning Wei, \textit{Member, IEEE}
		\thanks{Part of this article in~\cite{yang2025joint} has been presented in IEEE ICC workshop 2025. 
			
		Ran Yang, Yue Xiu, Wanting Lyu, Xiangxin Meng, Yan Li, and Ning Wei are with the National Key Laboratory of Wireless Communications, University of Electronic Science and Technology of China, Chengdu 611731, China (e-mail: yangran6710@outlook.com; xiuyue12345678@163.com; lyuwanting@yeah.net;   yan.li@uestc.edu.cn; wn@uestc.edu.cn).  
		
		Zheng Dong is with the School of Information Science and Engineering, Shandong University, Qingdao 266237, China (e-mail:  zhengdong@sdu.edu.cn). 
		
		Guangyi Liu is with the China Mobile Research Institute, Beijing 100053, China (email: liuguangyi@chinamobile.com).}
	}
	
	\markboth{Journal of \LaTeX\ Class Files,~Vol.~14, No.~8, August~2021}%
	{Shell \MakeLowercase{\textit{et al.}}: A Sample Article Using IEEEtran.cls for IEEE Journals}
	
	
	\maketitle
	\begin{abstract}
		Movable antennas (MAs) have demonstrated significant potential in enhancing the performance of dual-functional radar-communication (DFRC) systems. In this paper, we explore an MA-aided DFRC system that utilizes  a reconfigurable intelligent surface (RIS) to enhance signal coverage for communications in dead zones. To enhance the radar sensing performance in practical DFRC environments,  we propose a unified robust transceiver  design framework aimed at  maximizing the minimum  radar signal-to-interference-plus-noise ratio (SINR) in a cluttered environment. Our approach jointly optimizes transmit beamforming, receive filtering, antenna placement, and RIS reflecting coefficients under imperfect channel state information (CSI) for both sensing and communication channels. To deal with the  channel uncertainty-constrained issue,  we leverage the convex hull method to transform the primal problem into a more tractable form. We then introduce a two-layer block coordinate descent (BCD) algorithm, incorporating fractional programming (FP), successive convex approximation (SCA),  S-Lemma, and penalty techniques to reformulate it into a series of semidefinite program (SDP) subproblems that can be efficiently solved. We provide a comprehensive analysis of the convergence and computational complexity for the proposed design framework. Simulation results demonstrate the robustness of the proposed method, and show that the MA-based  design framework can significantly enhance the radar SINR performance while achieving  an effective balance between the radar and communication performance. 
	\end{abstract}
	
	\begin{IEEEkeywords}
		Movable antenna, dual-functional radar-communication, reconfigurable intelligent surface.
	\end{IEEEkeywords}

	\section{Introdution}
	\IEEEPARstart{A}{s} the landscape of wireless communications continues to evolve, the anticipation surrounding the advent of sixth generation (6G) networks is steadily growing\cite{rajatheva2020white,10753482}. According to the Federal Communications Commission (FCC), the upper mid-band spectrum resource (7-24 GHz) has recently been  identified with its opportunities to support various 6G usage scenarios including immersive communication, hyper reliable and low-latency communication (HRLLC), massive communication, artificial intelligence (AI) and communication, ubiquitous connectivity, and integrated sensing and communication (ISAC)\cite{10559933}. These usage scenarios form the vision of International Mobile
	Telecommunication (IMT)-2030, and necessitate a paradigm shift in current fifth generation (5G) wireless networks~\cite{saad2019vision}. In particular, dual-functional radar-communications (DFRC),
	which integrates sensing and communication functions into a single platform, has been envisioned as a key transformative technology for the upcoming
	generation~\cite{9737357}. DFRC systems share spectrum resources, hardware facilities, and signal-processing modules, resulting in significantly increased spectrum efficiency~\cite{2023arXiv230801227K}.  Thanks to these advantages, DFRC has shown great potential in various wireless systems, such as orthogonal time frequency space modulation (OTFS)\cite{yang2024sensing}, non-orthogonal multiple access (NOMA)\cite{lyu2023hybrid}, and unmanned aerial vehicle (UAV)\cite{wang2020constrained}.

	Despite extensive research on DFRC~\cite{yang2024sensing,lyu2023hybrid,wang2020constrained,10844869}, conventional {multiple-input multiple-output} (MIMO) systems with fixed-position antennas (FPAs) still present significant challenges in terms of power consumption and interference mitigation~\cite{gonzalez2025six}.  Specifically, FPA array configurations limit the full exploration  of diversity and spatial multiplexing gains of wireless channels, as they do not fully utilize channel variations across continuous spatial fields. Furthermore, fixed geometric configurations of antenna arrays  can lead to array gain loss during the radar beamforming tasks. Therefore, beyond developing advanced optimization algorithms and intelligent resource management strategies, introducing new revolutionary technologies such as reconfigurable antennas  is also critically important.
	
	{Recently, movable antennas (MAs), also named as fluid antennas\cite{9264694}, have been proposed as an innovative solution to overcome the fundamental limitations of conventional FPA-based systems~\cite{zhu2025tutorial}. In MA-assisted systems, each antenna element is connected to a radio frequency (RF) chain via flexible cables to support active movement, thereby reconstructing channel conditions to boost both the communication and radar performance. A prototype of the MA-assisted radar system was initially  demonstrated in \cite{zhuravlev2015experimental}, whereas the latest version demonstrates positioning accuracy of up to 0.05 mm~\cite{dong2024movable}. Mechanisms of antenna movement were investigated in \cite{11007274}. To reduce the hardware cost and complexity of conventional MA-based systems in~\cite{10286328}, a novel architecture named  cross-link MA array was proposed in~\cite{11226954}. Channel modeling and performance analysis were explored in \cite{zhu2023modeling}, where the authors established a field-response channel model under the far-field conditions. These foundational theoretical results provide a necessary framework for characterizing the performance limits and guiding the optimization of MA-enabled systems.}
	
{In recent years, several studies have demonstrated that the DFRC  performance can be significantly boosted by properly adjusting the positions of MAs within a designated area~\cite{lyu2025movable,wang2024fluid,jiang2025movable,hao2024fluid,xiu2025movable,guo2024movable}. For instance, in \cite{lyu2025movable}, the authors proposed a three-stage search-based projected gradient ascent (GA) method  to enhance the DFRC performance.  The {authors} in \cite{wang2024fluid} investigated a deep reinforcement learning method for the optimization on joint port selection and precoder for the MA-enhanced DFRC systems.
	In addition, the MA-based cooperative DFRC network was investigated in~\cite{xiu2025movable,guo2024movable}. To reduce the associated computational complexity and signaling overhead, the authors in \cite{10388242, 10734201} developed efficient optimization frameworks, including a gradient descent (GD) method tailored for zero-forcing (ZF) receivers and a two-timescale beamforming design that optimizes antenna placement based on statistical channel state information (CSI).
	Furthermore, with operations in higher frequency bands and larger antenna movement region, the authors in~\cite{10909572} extended the field-response channel model  from the far-field to the near-field propagation condition. Following this trajectory, the authors in~\cite{ding2024movable} pioneered investigation of MA-aided full-duplex DFRC systems under near-field channel conditions.  Although these studies  have demonstrated the superiority of MA over FPA, the susceptibility of  radio signals to blockage and attenuation events has not {yet} been fully investigated, which inevitably limits the practical  potential of the MA technique.}
	
	{Reconfigurable intelligent surface (RIS) is considered as a promising technology to address the aforementioned challenges\cite{pan2022overview}. A RIS is a planar array made up of many passive reflecting elements with controllable phase shifts\cite{wu2021intelligent}. Consequently, the RIS can not only reconfigure the wireless channels but also provide the line-of-sight (LoS) links for transmission {tasks} in dead zones. Recent studies have initially explored the integration of RIS into the MA-assisted DFRC systems\cite{ma2024movable,wu2025movable}. Specifically, an MA-assisted secure transmission scheme for the RIS-DFRC systems was investigated in \cite{ma2024movable}. The authors in \cite{wu2025movable} focused on maximizing the minimum beampattern gain for the MA-based RIS-DFRC systems. Nevertheless, these models are relatively simple, considering only clutter-free environments~\cite{ma2024movable} or focusing on designing the transmitter \cite{wu2025movable}, which inevitably leads to degraded communication and sensing performance. Furthermore, these works typically assume perfect knowledge of target locations and  communication channels, and the impact of sensing and communication channel errors has not been thoroughly studied yet. Although robust designs under imperfect CSI in DFRC systems \cite{liu2022outage,10443657,10497104} and RIS-aided systems \cite{10870062,9180053,10810484}, respectively, have been extensively investigated, existing approaches cannot be directly applied to MA-enhanced systems due to the complex  field-response channel model and unique spatial coupling between continuous antenna positions and channel uncertainty. As a result, to fully unlock the potential of MAs, a unified MA-based transceiver design framework to tackle these issues is greatly desirable. }

	In this paper, we investigate a robust  transceiver design for RIS-enhanced DFRC systems with movable antennas under both imperfect radar and communication channels. Specifically, we consider a monostatic DFRC system where a dual-functional base station (BS) equipped with 2D transceiver MA arrays serves multiple users and detects a point-like target in a cluttered environment.  A RIS is deployed to create virtual LoS links for the communications in dead zones.
	In particular, our contributions can be summarized as follows.
	%
	\begin{itemize}
		\item  To enhance radar sensing performance, a unified robust transceiver design  framework is developed to maximize the radar SINR  by jointly optimizing the positions of MAs, the transmitting beamforming, the receiving filter, and the RIS reflecting coefficients under imperfect CSI for both sensing and communication channels.	
		\item To address the channel uncertainty-constrained optimization problem,   we first leverage the convex hull method to transform the primal problem into a more tractable form. Consequently, we develop a two-layer block coordinate descent-based algorithm, combined with fractional programming (FP), successive convex approximation (SCA), S-Lemma, and penalty techniques to convert it into a sequence of semidefinite program (SDP) subproblems that can be efficiently solved.
		\item  To provide a performance benchmark, we extend the proposed framework to the perfect CSI case as an upper bound. We further provide a comprehensive analysis of the convergence and computational complexity for the proposed algorithms in both cases with perfect and imperfect CSI. 
		\item  
		Simulation results demonstrate that the proposed MA-based transceiver design framework exhibits remarkable robustness across various levels of channel uncertainty, and can significantly improve the minimum radar SINR while achieving  a satisfactory trade-off between the radar and communication performance. 
	\end{itemize} 
	
	\textit{Notations:} Boldface lower-case and upper-case letters indicate column vectors and matrices, respectively. $(\cdot)^*$, $(\cdot)^T$, $(\cdot)^H$, and $(\cdot)^{-1}$ denote the conjugate, transpose, transpose-conjugate, and inverse operations, respectively. $\mathbb{C}$ and $\mathbb{R}$ denote the sets of complex numbers and real numbers, respectively. $|a|$, $\|\boldsymbol{a}\|$, and $\|\boldsymbol{A}\|_F$ are the magnitude of a scalar $a$, the norm of a vector $\boldsymbol{a}$, and the Frobenius norm of a matrix $\boldsymbol{A}$, respectively. $\mathbb{E}\{\cdot\}$ represents statistical expectation. $\text{Tr}\{\boldsymbol{A}\}$ takes the trace of the matrix $\boldsymbol{A}$. $\Re\{\cdot\}$ and $\Im\{\cdot\}$ denote the real and imaginary parts of a complex number, respectively. $\angle a$ is the angle of complex-valued $a$. $\boldsymbol{I}_M$ indicates an $M \times M$ identity matrix.

	\section{System Model}
	\begin{figure}[t]
		\centerline{\includegraphics[width=3.0in]{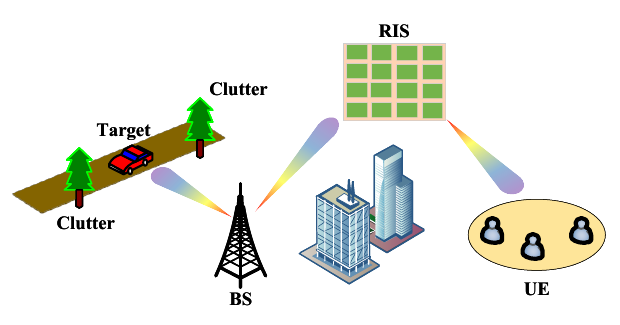}}
		\captionsetup{font={small},labelsep=period,singlelinecheck=off}
		\caption{The RIS-enhanced DFRC system with movable antennas.}
		\label{model}
	\end{figure}
	We consider a DFRC system as depicted in Fig.\,\ref{model}, where the dual-functional BS equipped with a planar array consisting of $N$ transmitting/receiving movable antennas serves $K$ single-antenna users while detecting a {point-like target} in the presence of $Q$ clutters. It is assumed that the direct links from the BS to the users are not available due to blockages, thus an $M$-element RIS is deployed to create virtual {LoS} links\footnote{{In this work, the RIS is deployed exclusively to establish virtual LoS links for communication, omitting its role in radar sensing. This is justified by the fact that the echo signals reflected via RIS undergo at least three-hop propagation (e.g., BS-RIS-Target-BS), resulting in significant product path loss. Consequently, their power is negligible compared to that of the direct sensing link. Nonetheless, the proposed algorithm can be readily adapted to scenarios involving RIS-assisted sensing.}}. The 2D moving regions for the transceiver antennas are respectively denoted by $\mathcal{C}_t$ and $\mathcal{C}_r$, both consisting of a square region of size $A\times A$. Moreover, the positions of the $n$-th transmitting and receiving antenna MAs are denoted by $\boldsymbol{t}_n = [x_n^t,y_n^t]^T$ and $\boldsymbol{r}_n = [x_n^r,y_n^r]^T$, with the reference points for the regions $\mathcal{C}_t$ and $\mathcal{C}_r$ represented by $\boldsymbol{o}^t = \boldsymbol{o}^r = [0,0]^T$, respectively.
	\subsection{Communication Model}
	Letting $\boldsymbol{w}_k$ represent the beamforming vector for the user $k$, the signal $\boldsymbol{x}\in \mathbb{C}^{N\times 1}$ transmitted by the BS can be written as
	\begin{equation}
		\boldsymbol{x} =\boldsymbol{Ws} = \sum_{k=1}^{K}\boldsymbol{w}_ks_k,
	\end{equation} 
	where $\boldsymbol{W} = [\boldsymbol{w}_1,\dots,\boldsymbol{w}_K]\in\mathbb{C}^{N\times K}$, and $\boldsymbol{s} = [s_1,\dots,s_K]^T\in \mathbb{C}^{K\times1}$ with $\mathbb{E}[\boldsymbol{s}\boldsymbol{s}^H] = \boldsymbol{I}_K$ represents the data symbols for all the users~\footnote{We  assmue that the BS generates the communication signals only to perform radar sensing, which requires least changes in existing wireless networks. Moreover, such DFRC waveforms should not bring more interference due to the absence of the dedicated radar signal.}. Given that the signal propagation distance is significantly larger than the size of the moving regions, the {far-field response} can be applied for channel modeling~\cite{10243545}. Specifically, the angle-of-arrival (AoA), angle-of-departure (AoD), and amplitude of the complex coefficient for each link remain constant despite the movement of the MAs. Note that we adopt the geometric model for the communication channels, thus the number of transmission paths at different nodes is the same, denoted by $L$\cite{zhu2023modeling}.  The elevation and azimuth angles of the $j$-th transmission path  at the BS and RIS are given by  $\psi^e_j \in [0,\pi],\psi^a_j\in [0,\pi]$ and $\phi^e_j \in [0,\pi],\phi^a_j\in [0,\pi]$, respectively. Then, for the $j$-th transmission path, the signal propagation difference between the position of the $n$-th transmitting MA $\boldsymbol{t}_n$ and the reference point $\boldsymbol{o}^t$ is given by
	\begin{equation}\label{rho}
		\rho(\boldsymbol{t}_n,\psi^e_j,\psi_j^a) = x^t_n\sin\psi^e_j\cos\psi^a_j+y^t_n\cos\psi^e_j.
	\end{equation} 
	Then, the field response vector (FRV) at  $\boldsymbol{t}_n$ can be given by
	\begin{equation}
		\boldsymbol{g}(\boldsymbol{t}_n) = \left[e^{\jmath\frac{2\pi}{\lambda}\rho(\boldsymbol{t}_n,\psi^e_1,\psi_1^a)},\dots,e^{\jmath\frac{2\pi}{\lambda}\rho(\boldsymbol{t}_n,\psi^e_{L},\psi_{L}^a)}\right]^T \in \mathbb{C}^{L \times 1},
	\end{equation} 
	where $\lambda$ is the carrier wavelength. Therefore,  the field response matrix (FRM) of the BS-RIS link for all $N$ transmitting MAs is given by
	\begin{equation}
		\boldsymbol{G}(\tilde{\boldsymbol{t}})\triangleq\left[\boldsymbol{g}(\boldsymbol{t}_1),\boldsymbol{g}(\boldsymbol{t}_2),\dots,\boldsymbol{g}(\boldsymbol{t}_N)\right]\in \mathbb{C}^{L\times N},
	\end{equation}
	where $\tilde{\boldsymbol{t}}\triangleq\left[\boldsymbol{t}_1^T,\boldsymbol{t}_2^T,\dots,\boldsymbol{t}_N^T\right]^T\in \mathbb{R}^{2N\times 1}$. Similarly, the FRV at the $m$-th RIS reflecting element can be derived as 
	\begin{equation}
		\boldsymbol{f}(\boldsymbol{s}_m) = \left[e^{\jmath\frac{2\pi}{\lambda}\rho(\boldsymbol{s}_m,\phi_1^e,\phi_1^a)},\dots,e^{\jmath\frac{2\pi}{\lambda}\rho(\boldsymbol{s}_m,\phi_{L}^e,\phi_{L}^a)}\right]^T\in \mathbb{C}^{L\times1},
	\end{equation}
	where $\boldsymbol{s}_m = [x_m^s,y_m^s]^T$ is the coordinate of the $m$-th element, and $\rho(\boldsymbol{s}_m,\phi_j^e,\phi_j^a)$ denotes the propagation distance difference  for the $j$-th path between the position $\boldsymbol{s}_m$ and origin of the RIS,  $\boldsymbol{o}^s = [0,0]^T$. Then, the FRM at the RIS can be given by
	\begin{equation}
		\boldsymbol{F}({\tilde{\boldsymbol{s}}}) \triangleq \left[\boldsymbol{f}(\boldsymbol{s}_1),\boldsymbol{f}(\boldsymbol{s}_2),\dots,\boldsymbol{f}(\boldsymbol{s}_M)\right]^T\in \mathbb{C}^{L\times M}.
	\end{equation}
	Let $\boldsymbol{\Sigma}=\mathrm{diag}\{\sigma_{1,1},\sigma_{2,2},\dots,\sigma_{L,L}\}\in \mathbb{C}^{L\times L}$ denote the path response matrix (PRM) of the BS-RIS link, the channel matrix can be expressed as 
	\begin{equation}\label{channel}
		\boldsymbol{H}(\tilde{\boldsymbol{t}}) = \boldsymbol{F}({\tilde{\boldsymbol{s}}})^H\boldsymbol{\Sigma}\boldsymbol{G}(\tilde{\boldsymbol{t}})\in \mathbb{C}^{M\times N}.
	\end{equation}
	The  channel $\boldsymbol{h}_k\in \mathbb{C}^{M\times 1}$ between the RIS and the $k$-th user can be derived in a similar fashion as (\ref{channel}) and is thus omitted for brevity. The received signal of the $k$-th user is given by
	\begin{equation}
		y_k = \underbrace{\boldsymbol{h}_k^H\boldsymbol{V}\boldsymbol{H}(\tilde{\boldsymbol{t}})\boldsymbol{w}_ks_k}_{\text{desired signal}}+\underbrace{\sum_{j=1,j\neq k}^{K}\boldsymbol{h}_k^H\boldsymbol{V}\boldsymbol{H}(\tilde{\boldsymbol{t}})\boldsymbol{w}_js_j}_{\text{inter-user interference}}+n_k,
	\end{equation}
	where $\boldsymbol{V} = \mathrm{diag}\{[v_1,\dots,v_M]^T\}\in \mathbb{C}^{M\times M}$ with $v_m = e^{\jmath\theta_m}$ {being} the reflection coefficient matrix, and $n_k \sim \mathcal{CN}(0,\sigma_k^2)$ is the additive white Gaussian noise (AWGN).  
	Then, the SINR of the $k$-th user is given by
	\begin{equation}
		\Gamma_k(\boldsymbol{W},\boldsymbol{V},\tilde{\boldsymbol{t}}) = \frac{|\boldsymbol{h}_k^H\boldsymbol{V}\boldsymbol{H}(\tilde{\boldsymbol{t}})\boldsymbol{w}_k|^2}{\sum_{j=1,j\neq k}^{K}|\boldsymbol{h}_k^H\boldsymbol{V}\boldsymbol{H}(\tilde{\boldsymbol{t}})\boldsymbol{w}_j|^2+\sigma_k^2}.
	\end{equation}
	\subsection{Radar Model}
	{We adopt the LoS channel model for the sensing channels between the BS and the target/clutters. Let $\varphi_0^e$ and $\varphi_0^a$ denote the elevation and azimuth angle between the target and the BS, the receiving and transmitting steering vectors can be given by $	\mathbf{a}_0^r(\varphi_0^e,\varphi_0^a,\tilde{\boldsymbol{r}}) = [e^{\jmath\frac{2\pi}{\lambda}\rho(\boldsymbol{r}_1,\varphi_0^e,\varphi_0^a)},\dots,e^{\jmath\frac{2\pi}{\lambda}\rho(\boldsymbol{r}_N,\varphi_0^e,\varphi_0^a)}]^T$  and $\mathbf{a}_0^t(\varphi_0^e,\varphi_0^a,\tilde{\boldsymbol{t}}) = [e^{\jmath\frac{2\pi}{\lambda}\rho(\boldsymbol{t}_1,\varphi_0^e,\varphi_0^a)},\dots,e^{\jmath\frac{2\pi}{\lambda}\rho(\boldsymbol{t}_N,\varphi_0^e,\varphi_0^a)}]^T$, respectively. Let  $\boldsymbol{A}_0(\tilde{\boldsymbol{r}},\tilde{\boldsymbol{t}}) = 	\mathbf{a}_0^r(\varphi_0^e,\varphi_0^a,\tilde{\boldsymbol{r}})\mathbf{a}_0^t(\varphi_0^e,\varphi_0^a,\tilde{\boldsymbol{t}})^H$ denote the response matrix for the target, and the received echo signal can be given by
	\begin{equation}
		y_r = \alpha_0 \boldsymbol{u}^H\boldsymbol{A}_0(\tilde{\boldsymbol{r}},\tilde{\boldsymbol{t}})\boldsymbol{x} + \boldsymbol{u}^H\boldsymbol{c} + \boldsymbol{u}^H\boldsymbol{n}_r,		
	\end{equation}
	where  $\boldsymbol{c} = \sum_{q=1}^{Q}\alpha_q\boldsymbol{A}_q(\tilde{\boldsymbol{r}},\tilde{\boldsymbol{t}})\boldsymbol{x}$, $ \boldsymbol{A}_q(\tilde{\boldsymbol{r}},\tilde{\boldsymbol{t}}) =  \mathbf{a}_q^r(\varphi_q^e,\varphi_q^a,\tilde{\boldsymbol{r}})\mathbf{a}_q^t(\varphi_q^e,\varphi_q^a,\tilde{\boldsymbol{t}})^H$ denotes the response matrix for the $q$-th clutter, and $\boldsymbol{u}\in \mathbb{C}^{N\times1}$  is the receiving filter\footnote{Generally, the clutter interference can be modeled to be signal-independent or signal-dependent. We consider the signal-dependent clutter interference in this article,  and our method can be directly applied to deal with the signal-independent interference.}.  Here, $\alpha_0$ and $\alpha_q$ are the complex coefficients including the  radar cross section (RCS) and the cascaded complex gain of the target and $q$-th clutter, with $\mathbb{E}\{|\alpha_0|^2\} = \zeta_0^2 ~\text{and}~ \mathbb{E}\{|\alpha_q|^2\} = \zeta_q^2$, respectively. Besides, $\boldsymbol{n}_r \sim\mathcal{CN}(0,\sigma_r^2\boldsymbol{I}_N)$ is the AWGN at the BS. {We assume that self-interference (SI) between the transmit and receive antenna arrays has been effectively suppressed by advanced SI cancellation methods, such as physical isolation and digital cancellation~\cite{8642523,11157898,11079638}}\footnote{{The proposed optimization approach can be directly adapted to support active suppression of SI.}}. According to~\cite{hao2024fluid}, the radar detection probability monotonically increases with  sensing
	SINR. Therefore, to improve the detection probability, we adopt sensing SINR as a radar performance metric. Specifically,  the instantaneous radar SINR is given by
		\begin{equation}
			{\Gamma}_I (\boldsymbol{W},\tilde{\boldsymbol{r}},\tilde{\boldsymbol{t}},\boldsymbol{u}) = \frac{|\alpha_0 \boldsymbol{u}^H\boldsymbol{A}_0(\tilde{\boldsymbol{r}},\tilde{\boldsymbol{t}})\boldsymbol{x}|^2}{\boldsymbol{u}^H(\boldsymbol{c}\boldsymbol{c}^H+\boldsymbol{n}_r\boldsymbol{n}_r^H)\boldsymbol{u}}.
	\end{equation}} To maximize the radar SINR, the optimal solution for the filter $\boldsymbol{u}^\star$ can be obtained by solving the minimum variance distortionless response (MVDR) problem~\cite{yang2025joint}, i.e., 
	\begin{equation}\label{filter}
		\boldsymbol{u}^\star = \beta\big(\boldsymbol{c}\boldsymbol{c}^H+\boldsymbol{n}_r\boldsymbol{n}_r^H\big)^{-1}\boldsymbol{A}_0(\tilde{\boldsymbol{r}},\tilde{\boldsymbol{t}})\boldsymbol{x},
	\end{equation}
	where $\beta$ is an auxiliary constant. Then, the average radar SINR can be derived as~\cite{wang2024fluid}, i.e., 
	\begin{align} \label{SINR}
		&\Gamma_A(\boldsymbol{W},\tilde{\boldsymbol{r}},\tilde{\boldsymbol{t}})\notag\\ &\mathop = \limits ^{\left ({a}\right)}~
		{\mathbb{E}}\left[ |\alpha_0|^2\boldsymbol{x}^H\boldsymbol{A}_0(\tilde{\boldsymbol{r}},\tilde{\boldsymbol{t}})^H\big(\boldsymbol{c}\boldsymbol{c}^H+\boldsymbol{n}_r\boldsymbol{n}_r^H\big)^{-1}\boldsymbol{A}_0(\tilde{\boldsymbol{r}},\tilde{\boldsymbol{t}})\boldsymbol{x}\right] \notag \\
		&\mathop \ge \limits ^{\left ({b }\right)}~ {\mathrm{tr}}\left({\boldsymbol{\Phi}}\boldsymbol{W}\boldsymbol{W}^H\right) \triangleq \Gamma_r(\boldsymbol{W},\tilde{\boldsymbol{r}},\tilde{\boldsymbol{t}}),
	\end{align} 
	where ${\boldsymbol{\Phi}} = \zeta_0^2\boldsymbol{A}_0(\tilde{\boldsymbol{r}},\tilde{\boldsymbol{t}})^H(\boldsymbol{\Xi}+\sigma_r^2\boldsymbol{I}_N)^{-1}\boldsymbol{A}_0(\tilde{\boldsymbol{r}},\tilde{\boldsymbol{t}}),$ and  $\boldsymbol{\Xi} = \sum_{q=1}^{Q}\zeta_q^2\boldsymbol{A}_q(\tilde{\boldsymbol{r}},\tilde{\boldsymbol{t}})\boldsymbol{W}\boldsymbol{W}^H\boldsymbol{A}_q(\tilde{\boldsymbol{r}},\tilde{\boldsymbol{t}})^H$. Note that the procedure (a) is due to the optimal receiving filter $\boldsymbol{u}^\star$ in \eqref{filter}, and the procedure (b) holds due to Jensen's inequality, $\mathbb{E}[ \boldsymbol{xx}^H] =  \boldsymbol{W}\boldsymbol{W}^H$,  $\mathbb{E}[ \boldsymbol{cc}^H] = \boldsymbol{\Xi}$, and $\mathbb{E}\{|\alpha_0|^2\} = \zeta_0^2$.
	\subsection{Channel Uncertainty Model}
	In  practical DFRC scenarios, acquiring perfect knowledge for both communication and sensing channels is  challenging. Specifically, for the communication part, although various channel estimation methods have been developed for the MA and RIS-aided communications \cite{pan2022overview,zhu2025tutorial}, the perfect CSI for the reflecting links is difficult to obtain due to the limited signal processing capabilities of passive RIS elements, and fast change of user locations. To characterize this effect, we exploit the bounded CSI error model to quantify the CSI uncertainty of the RIS-user channels~\cite{5164911}, i.e.,
	\begin{equation}
		\boldsymbol{h}_k = \hat{\boldsymbol{h}}_k + \Delta \boldsymbol{h}_k, ||\Delta \boldsymbol{h}_k||_2\le \varepsilon_k,
	\end{equation}
	where $\hat{\boldsymbol{h}}_k$ is the estimated channel vector, and $\varepsilon_k$ is the radius of the uncertainty region for the $k$-th user known by the BS. 
	
	For the sensing part, we assume that the target location is not perfectly known to the BS due to its movement and random fluctuation~\cite{su2022secure}. That is, the target is located in an uncertain angular interval, such that
	
	\begin{subequations} \label{angleregion}
		\begin{alignat}{2}
			&\Psi_e = \big[\vartheta_e-\frac{1}{2}\Delta\vartheta_e,\vartheta_e+\frac{1}{2}\Delta\vartheta_e\big],\\
			&	\Psi_a = \big[\vartheta_a-\frac{1}{2}\Delta\vartheta_a,\vartheta_a+\frac{1}{2}\Delta\vartheta_a\big],
		\end{alignat}
	\end{subequations}
	where $\Delta\vartheta_e$ and $\Delta\vartheta_a$ denote the elevation and azimuth error range, respectively.

	\subsection{Problem Formulation}
	Based on the above performance metrics, we focus on the robust transceiver design under both imperfect communication and sensing channels. Specifically, we aim to maximize the minimum radar SINR performance with regard to {worst-case} locations within the uncertain interval $\Psi_e$ and $\Psi_a$ subject to the given QoS constraints under the bounded CSI error model. This is achieved by jointly designing the transmitting beamforming vectors, positions of transceiver antennas, and RIS reflecting elements. In particular, the optimization problem can be formulated as 
	\begin{subequations}\label{robust1}
		\begin{alignat}{2}
			&\underset{\boldsymbol{W},\boldsymbol{V},\tilde{\boldsymbol{t}},\tilde{\boldsymbol{r}}}{\max} ~~\underset{\substack{\varphi_0^e\in\Psi_e\\\varphi_0^a\in\Psi_a}}{\min} ~~\Gamma _r(\boldsymbol{W},\tilde{\boldsymbol{r}},\tilde{\boldsymbol{t}})\label{robust1_objective}\\
			&\,\text {s.t.}~ \Gamma_k(\boldsymbol{W},\boldsymbol{V},\tilde{\boldsymbol{t}}) \ge \gamma_k,||\Delta\boldsymbol{h}_k||_2\le\varepsilon_k,\forall k,\label{robust1_Com}\\
			&\hphantom {s.t.~}||\boldsymbol{t}_n-\boldsymbol{t}_{n'}||_2^2 \ge D^2,|| \boldsymbol{r}_n-\boldsymbol{r}_{n'}||_2^2 \ge D^2,\forall n\neq n',\label{robust1_MA1}\\
			&\hphantom {s.t.~}  \boldsymbol{t}_n \in \mathcal{C}_t,\boldsymbol{r}_n \in \mathcal{C}_r,\forall n,\label{robust1_MA2}\\
			&\hphantom {s.t.~}
			\sum_{k=1}^{K}\boldsymbol{w}_k^H\boldsymbol{w}_k \le P_t,\label{robust1_Pt}\\
			&\hphantom {s.t.~}
			|{v}_m|^2 = 1, \forall m,\label{robust1_RIS}	
		\end{alignat} 
	\end{subequations} 
	where the constraints in \eqref{robust1_Com} ensure that the SINR at the $k$-th user is no less than the predefined threshold $\gamma_k$, $D$ represents the minimum distance between the MAs to prevent  coupling effects, $P_t$ is the maximum transmission power, and  \eqref{robust1_RIS} are the unit-modulus constraints on RIS.  It is challenging to solve the above problem due to the non-convexity of~\eqref{robust1_Com},~\eqref{robust1_MA1},~\eqref{robust1_MA2},~\eqref{robust1_Pt}, as well as the coupling between optimization  variables. The angle uncertainty and CSI errors further increase the difficulty of solving the problem in~\eqref{robust1}.

	
	\section{Two-layer BCD Algorithm}
	Note that the inaccurate location region of the sensing target  is a continuous area, which makes the problem in~\eqref{robust1}   non-convex and  mathematically intractable. The random communication CSI errors further increase the difficulty. To deal with these issues, we leverage the convex hull technique to  transform the original problem in~\eqref{robust1} into a more tractable form, and develop a two-layer  block coordinate descent (BCD) algorithm to solve it, the details of which are elaborated as follows.
	\subsection{Problem Transformation}
	Based on the concept of convex hull, any channel matrix in an uncertainty set can be represented as a weighted combination of discrete samples\cite{boyd2004convex}. Following this idea, we first uniformly discretize the continuous region,  with the number of sampling points denoted by $P_s$. Let $(\varphi_p^e,\varphi_p^a)$ denote the $p$-th possible location, and we construct a convex hull based on the weighted sum of $P_s$ steering matrix samples, i.e., 
	\begin{equation}
		\Upsilon = \Big\{\sum_{p=1}^{P_s}\mu_{p}\boldsymbol{A}_p(\tilde{\boldsymbol{r}},\tilde{\boldsymbol{t}})|\sum_{p=1}^{P_s}\mu_p=1,\mu_p\ge0\Big\},
	\end{equation}
	where $\boldsymbol{A}_p(\tilde{\boldsymbol{r}},\tilde{\boldsymbol{t}}) = \mathbf{a}_p^r(\varphi_p^e,\varphi_p^a,\tilde{\boldsymbol{r}})\mathbf{a}_p^t(\varphi_p^e,\varphi_p^a,\tilde{\boldsymbol{t}})^H$, and $\boldsymbol{\mu} = [\mu_1,\dots,\mu_{P_s}]^T\in \mathbb{C}^{P_s\times 1}$ is the weighted coefficient vector. Let  ${\tilde{\boldsymbol{\Phi}}} = \zeta_0^2\sum_{p=1}^{P_s}\mu_{p}\boldsymbol{A}_p(\tilde{\boldsymbol{r}},\tilde{\boldsymbol{t}})^H(\boldsymbol{\Xi}+\sigma_r^2\boldsymbol{I}_N)^{-1}\sum_{p=1}^{P_s}\mu_{p}\boldsymbol{A}_p(\tilde{\boldsymbol{r}},\tilde{\boldsymbol{t}})$, and we construct a surrogate function for (\ref{robust1_objective}) to facilitate the algorithm development, i.e., 
	\begin{equation}\label{robust1_surrogate}
		\tilde{\Gamma}_r(\boldsymbol{W},\tilde{\boldsymbol{r}},\tilde{\boldsymbol{t}},\boldsymbol{\mu}) = {\mathrm{tr}}\left({\tilde{\boldsymbol{\Phi}}}\boldsymbol{W}\boldsymbol{W}^H\right).
	\end{equation} 
	We then demonstrate  the relationship between (\ref{robust1_objective}) and (\ref{robust1_surrogate}), i.e., 
	\begin{align}
		&\underset{\boldsymbol{W},\boldsymbol{V},\tilde{\boldsymbol{t}},\tilde{\boldsymbol{r}}}{\max} ~\underset{\substack{\varphi_0^e\in\Psi_e\\\varphi_0^a\in\Psi_a}}{\min} ~\Gamma _r(\boldsymbol{W},\tilde{\boldsymbol{r}},\tilde{\boldsymbol{t}}) \ge\underset{\boldsymbol{W},\boldsymbol{V},\tilde{\boldsymbol{t}},\tilde{\boldsymbol{r}}}{\max} \underset{\boldsymbol{\mu}}{\min}~\tilde{\Gamma}_r(\boldsymbol{W},\tilde{\boldsymbol{r}},\tilde{\boldsymbol{t}},\boldsymbol{\mu}) 
		\notag\\  &=\underset{\boldsymbol{\mu}}{\min}\underset{\boldsymbol{W},\boldsymbol{V},\tilde{\boldsymbol{t}},\tilde{\boldsymbol{r}}}{\max}~ \tilde{\Gamma}_r(\boldsymbol{W},\tilde{\boldsymbol{r}},\tilde{\boldsymbol{t}},\boldsymbol{\mu}). 
	\end{align}
	
	\textit{Proof:} Please refer to the Appendix A in \cite{8334240}. \hfill$\Box$
	
	%
	It can be observed that the surrogate objective function in (\ref{robust1_surrogate}) is highly intractable due to the coupling of the optimization variables. To tackle this challenge, we decouple the revised problem into the following two parts, i.e.,
	\begin{align}
		&(\text {P1}):~~\max \limits _{\boldsymbol{W},\boldsymbol{V},\tilde{\boldsymbol{t}},\tilde{\boldsymbol{r}}}  \tilde{\Gamma}_r(\boldsymbol{W},\tilde{\boldsymbol{r}},\tilde{\boldsymbol{t}})\notag\\&\qquad \quad \,\text{~s.t}.~ \text{(\ref{robust1_Com}),~(\ref{robust1_MA1}),~(\ref{robust1_MA2}),~(\ref{robust1_Pt}),~(\ref{robust1_RIS})},\label{robust_P1}\\
		&(\text {P2}):~~\min \limits _{\boldsymbol {\mu}}~ \tilde{\Gamma}_r(\boldsymbol{\mu})\notag \\&\qquad \quad \, ~\text{s.t.~} \sum_{p=1}^{P_s}\mu_p=1,~\mu_p>=0.\label{robust_P2}
	\end{align}
	Then, we propose a two-layer BCD-based algorithm to solve the problems (P1) and (P2) in an iterative manner until convergence, the details of which are given as follows.
	\subsection{Optimization on (P1)}	
	To facilitate the algorithm development, we first employ the fractional programming method in ~\cite{8314727} to equivalently convert the objective function in~\eqref{robust_P1} into a more tractable form as
	\begin{align}\label{robust1_surrogate_reformulated}
		\bar{\Gamma}&_r(\boldsymbol{W},\tilde{\boldsymbol{r}},\tilde{\boldsymbol{t}},\boldsymbol{\Lambda}) = \notag\\
		&\zeta_0^2 \mathrm{tr}\left(2\Re\{\boldsymbol{W}^H\sum_{p=1}^{P_s}\mu_p\boldsymbol{A}_p(\tilde{\boldsymbol{r}},\tilde{\boldsymbol{t}})^H\boldsymbol{\Lambda}\}-\boldsymbol{\Lambda}^H(\boldsymbol{\Xi}+\sigma_r^2\boldsymbol{I}_N)\boldsymbol{\Lambda}\right),
	\end{align}
	where $\boldsymbol{\Lambda} \in \mathbb{C}^{N\times K}$ is an auxiliary variable. Based on the reformulated objective function in~\eqref{robust1_surrogate_reformulated}, we propose a BCD-based algorithm by incorporating SCA, S-Lemma, and penalty techniques  to solve the problem in~\eqref{robust_P1}, the  procedures of which are given as follows.
	\subsubsection{Updating Auxiliary Variable}
	Note that determining the optimal value of $\boldsymbol{\Lambda}$ is an unconstrained optimization problem, and the optimal solution  $\boldsymbol{\Lambda}^\star$ can be directly given by
	\begin{equation}\label{robust_Lambda}
		\boldsymbol{\Lambda}^\star =\left (\boldsymbol{\Xi}+\sigma_r^2\boldsymbol{I}_N\right)^{-1}\sum_{p=1}^{P_s}\mu_{p}\boldsymbol{A}_p(\tilde{\boldsymbol{r}},\tilde{\boldsymbol{t}})\boldsymbol{W}.
	\end{equation}
	
	\subsubsection{Updating Transmit Beamforming}
	With $\tilde{\boldsymbol{r}},\tilde{\boldsymbol{t}}, \boldsymbol{\Lambda}$ and $\boldsymbol{V}$ being fixed, the problem in~\eqref{robust_P1} can be recast as
	\begin{subequations}\label{robust1_beamforming1}
		\begin{alignat}{2}
			&\underset{ \boldsymbol{W}}{\max} \quad \bar{\Gamma}_r(\boldsymbol{W})\\
			&\,\text {s.t.}~\Gamma_k(\boldsymbol{W}) \ge \gamma_k,||\Delta\boldsymbol{h}_k||_2\le\varepsilon_k,\forall k,\label{robust1_beamforming1_Com}\\
			&\hphantom {s.t.~}
			\sum_{k=1}^{K}\boldsymbol{w}_k^H\boldsymbol{w}_k \le P_t.\label{robust1_beamforming1_Pt}
		\end{alignat} 
	\end{subequations} 
	It is observed that the non-convexity of the above problem lies in the constraints in  (\ref{robust1_beamforming1_Com}) over the CSI uncertainty regions. To deal with this issue, we rewrite the constraints in (\ref{robust1_beamforming1_Com}) as
	\begin{subequations}\label{sw}
		\begin{alignat}{2}
			&|\boldsymbol{h}_k^H\boldsymbol{V}\boldsymbol{H}(\tilde{\boldsymbol{t}})\boldsymbol{w}_k|^2\ge z_k\gamma_k, ||\Delta\boldsymbol{h}_k||_2 \le \varepsilon_k,\forall k,\label{robust1_beamforming1_Com11}\\
			&||\boldsymbol{h}_k^H\boldsymbol{V}\boldsymbol{H}(\tilde{\boldsymbol{t}})\boldsymbol{W}_{-k}||_2^2+\sigma_k^2\le z_k,||\Delta\boldsymbol{h}_k||_2 \le \varepsilon_k,\forall k,\label{robust1_beamforming1_Com22}
		\end{alignat} 
	\end{subequations}
	where $\boldsymbol{W}_{-k} = [\boldsymbol{w}_1,\dots,\boldsymbol{w}_{k-1},\boldsymbol{w}_{k+1},\dots,\boldsymbol{w}_{K}]\in \mathbb{C}^{N\times(K-1)},$ and $\boldsymbol{z} = [z_1,\dots,z_K]^T\in \mathbb{C}^{K\times1}$ are the auxiliary variables. The equivalence between~\eqref{sw} and~\eqref{robust1_beamforming1_Com} can be verified by
	contradiction. We first handle the non-convex quadratic inequalities in  (\ref{robust1_beamforming1_Com11}). Specifically, the left-hand-side (LHS) of (\ref{robust1_beamforming1_Com11}) is approximated with its lower bound, as shown below.
	
	\textit{Lemma 1:} Let  $\boldsymbol{w}_k^{(l)}$ and $\boldsymbol{V}^{(l)}$ be the optimal solutions obtained at the $l$-th iteration, then a linear lower bound of $|\boldsymbol{h}_k^H\boldsymbol{V}\boldsymbol{H}(\tilde{\boldsymbol{t}})\boldsymbol{w}_k|^2$ in (\ref{robust1_beamforming1_Com11}) at $(\boldsymbol{w}_k^{(l)},\boldsymbol{V}^{(l)})$ is given by
	\begin{equation}
		|\boldsymbol{h}_k^H\boldsymbol{V}\boldsymbol{H}(\tilde{\boldsymbol{t}})\boldsymbol{w}_k|^2\ge \boldsymbol{h}_k^H\boldsymbol{X}_k\boldsymbol{h}_k,\forall k,\label{robust1_beamforming1_Com11_lower1}
	\end{equation}
	where  
	\begin{align}
		\boldsymbol{X}_k =  \boldsymbol{V}\boldsymbol{H}&(\tilde{\boldsymbol{t}})\boldsymbol{w}_k\boldsymbol{w}_k^{H,(l)}\boldsymbol{H}(\tilde{\boldsymbol{t}})^H\boldsymbol{V}^{H,(l)}\notag\\
		&+\boldsymbol{V}^{(l)}\boldsymbol{H}(\tilde{\boldsymbol{t}})\boldsymbol{w}_k^{(l)}\boldsymbol{w}_k^H\boldsymbol{H}(\tilde{\boldsymbol{t}})^H\boldsymbol{V}^H\notag\\
		&-\boldsymbol{V}^{(l)}\boldsymbol{H}(\tilde{\boldsymbol{t}})\boldsymbol{w}_k^{(l)}\boldsymbol{w}_k^{H,(l)}\boldsymbol{H}(\tilde{\boldsymbol{t}})^H\boldsymbol{V}^{H,(l)},\notag
	\end{align} 
	and $\boldsymbol{V} = \boldsymbol{V}^{(l)}$ is treated as a constant term here.
	

	\textit{Proof:} Let $a$ be a complex scalar variable. By applying the Appendix B in \cite{8579566}, we have the inequality 
	\begin{equation*}
		|a|^2\ge a^{*,(l)}a + a^{*}a^{(l)}-a^{*,(l)}a^{(l)},
	\end{equation*}
	for any fixed $a^{(l)}$. Then, (\ref{robust1_beamforming1_Com11_lower1}) is obtained by replacing $a$ and $a^{(l)}$ with $\boldsymbol{h}_k^H\boldsymbol{V}\boldsymbol{H}(\tilde{\boldsymbol{t}})\boldsymbol{w}_k$ and $\boldsymbol{h}_k^H\boldsymbol{V}^{(l)}\boldsymbol{H}(\tilde{\boldsymbol{t}})\boldsymbol{w}_k^{(l)}$, respectively. The proof is completed. \hfill$\Box$
	
	With the aid of $\boldsymbol{h}_k = \hat{\boldsymbol{h}}_k + \Delta\boldsymbol{h}_k$ and \textit{Lemma 1}, the inequality in  (\ref{robust1_beamforming1_Com11}) is reformulated as
	\begin{equation}
		\Delta\boldsymbol{h}_k^H\boldsymbol{X}_k\Delta\boldsymbol{h}_k+2\Re\{\hat{\boldsymbol{h}}_k^H\boldsymbol{X}_k\Delta\boldsymbol{h}_k\}+\hat{\boldsymbol{h}}_k^H\boldsymbol{X}_k\hat{\boldsymbol{h}}_k\ge z_k\gamma_k,\forall k.\label{robust1_beamforming1_Com11_lower2}
	\end{equation}
	
	In order to tackle the CSI uncertainties  $\{\Delta\boldsymbol{h}_k\}_{k=1}^K$ in  (\ref{robust1_beamforming1_Com11}), the S-Lemma in \cite{9180053} is used to transform (\ref{robust1_beamforming1_Com11_lower2}) into a sequence of linear matrix inequalities (LMIs) as
	\begin{align} \label{robust1_beamforming1_Com11_lower3}
		\setlength{\arraycolsep}{0.2pt}
		\begin{bmatrix} 
			\varpi _{k}\boldsymbol {I}_{M}+\boldsymbol{X}_{k} & \boldsymbol{X}_k^H\hat{\boldsymbol{h}}_k \\ 
			\hat{\boldsymbol{h}}_k^H\boldsymbol{X}_k &\hat{\boldsymbol{h}}_k^H\boldsymbol{X}_k\hat{\boldsymbol{h}}_k - z_k\gamma_k-\varpi _{k}\varepsilon_k^2
		\end{bmatrix} \succeq \mathbf {0},\,\forall k, 
	\end{align} 
	where $\boldsymbol{\varpi} = [\varpi_1,\dots,\varpi_K]^T\ge 0$ are the slack variables. 
	
	For the constraints in  (\ref{robust1_beamforming1_Com22}), we first adopt the Schur's complement in~\cite{boyd2004convex} to recast the inequalities in (\ref{robust1_beamforming1_Com22}) as 
	\begin{align} \label{robust1_beamforming1_Com22_lower1}
		\setlength{\arraycolsep}{0.2pt}
		\begin{bmatrix} 
			z _{k}-\sigma _{k}^{2} & \boldsymbol{h}_k^H\boldsymbol{VH}(\tilde{\boldsymbol{t}})\boldsymbol{W}_{-k}\\ 
			(\boldsymbol{h}_k^H\boldsymbol{VH}(\tilde{\boldsymbol{t}})\boldsymbol{W}_{-k})^H &\boldsymbol{I}_M
		\end{bmatrix} \succeq \mathbf {0},\,\forall k, 
	\end{align} 
	Then, by using the Nemirovski's lemma in \cite{1369660} and introducing the slack variables $\boldsymbol{\xi} = [\xi_1,\dots,\xi_K]^T\ge 0$, the LMIs in~(\ref{robust1_beamforming1_Com22_lower1}) can be further rewritten as 
	\begin{small}
		\begin{align} \label{robust1_beamforming1_Com22_lower2}
			\setlength{\arraycolsep}{0.5pt}
			\begin{bmatrix} 
				z_k-\sigma_k^2-\xi_k & \hat{\boldsymbol{h}}_k^H\boldsymbol{VH}(\tilde{\boldsymbol{t}})\boldsymbol{W}_{-k} & \boldsymbol{0}_{1\times M}\\ 
				(\hat{\boldsymbol{h}}_k^H\boldsymbol{VH}(\tilde{\boldsymbol{t}})\boldsymbol{W}_{-k})^H&\boldsymbol{I}_{K-1}  &\varepsilon_k(\boldsymbol{V}\boldsymbol{H}(\tilde{\boldsymbol{t}})\boldsymbol{W}_{-k})^H\\
				\boldsymbol{0}_{M\times1} & \varepsilon_k\boldsymbol{V}\boldsymbol{H}(\tilde{\boldsymbol{t}})\boldsymbol{W}_{-k} & \xi_k\boldsymbol{I}_M
			\end{bmatrix} \notag\\
			\succeq \mathbf {0},\,\forall k. 
		\end{align} 
	\end{small}
	
	Combining the LMIs in (\ref{robust1_beamforming1_Com11_lower3}) and
	(\ref{robust1_beamforming1_Com22_lower2}), the problem in  (\ref{robust1_beamforming1}) can be approximated as 
	\begin{subequations}\label{robust1_beamforming2}
		\begin{alignat}{2}
			&\underset{ \boldsymbol{W},\boldsymbol{\varpi},\boldsymbol{\xi},\boldsymbol{z}}{\max} ~ \bar{\Gamma}_r(\boldsymbol{W})\\
			&\,\text {s.t.}~\text{(\ref{robust1_beamforming1_Pt}),~(\ref{robust1_beamforming1_Com11_lower3}),~(\ref{robust1_beamforming1_Com22_lower2})},\boldsymbol{\varpi}\ge 0, \boldsymbol{\xi}\ge 0,
		\end{alignat} 
	\end{subequations} 
	which is a semidefinite program and can be solved by using the CVX tool~\cite{boyd2004convex}.
	\subsubsection{Updating Reflecting Coefficients}
	Now, we carry out optimization on RIS reflecting  coefficients with $\boldsymbol{W},\tilde{\boldsymbol{r}},\tilde{\boldsymbol{t}}$, and $\boldsymbol{\Lambda}$ being fixed.  Note that the objective function  $\bar{\Gamma}_r(\boldsymbol{W},\tilde{\boldsymbol{r}},\tilde{\boldsymbol{t}},\boldsymbol{\Lambda})$ is independent of $\boldsymbol{V}$, which indicates that the design of $\boldsymbol{V}$ is a feasibility-check problem and the solution will not directly affect  \eqref{robust1_objective}. Therefore, in order to provide additional degrees of freedom (DoFs) for optimization on other variables, we propose to maximize the minimum user SINR by updating $\boldsymbol{V}$~\cite{pan2022overview}. Specifically, we introduce a slack variable $\eta$ to recast the problem (P1)  in~\eqref{robust_P1} as 
	\begin{subequations}\label{robust_RIS}
		\begin{alignat}{2}
			&\underset{ \boldsymbol{V},\eta}{\max} \quad \eta\label{robust_RIS_OB}\\
			&\,\text {s.t.}~ \Gamma_k(\boldsymbol{V})\ge \eta,||\Delta\boldsymbol{h}_k||_2\le \varepsilon_k,\forall k,\label{robust_RIS_Com}\\
			&\hphantom {s.t.~}
			|{v}_m|^2 = 1,\forall m.\label{robust_RIS_Unit}	
		\end{alignat} 
	\end{subequations} 
	The problem in~\eqref{robust_RIS} is intractable due to the non-convexity of~\eqref{robust_RIS_Com} and~\eqref{robust_RIS_Unit}. We first deal with the constraints in~\eqref{robust_RIS_Com}. Following the similar steps in updating transmitting beamforming, we arrive at the following problem as
	\begin{subequations}\label{robust1_RIS1}
		\begin{alignat}{2}
			&\underset{ \boldsymbol{V},\boldsymbol{\varpi},\boldsymbol{\xi},\boldsymbol{z},\eta}{\max} ~ \eta\\
			&\,\text {s.t.}~\text{Modified-(\ref{robust1_beamforming1_Com11_lower3}),~(\ref{robust1_beamforming1_Com22_lower2})},\boldsymbol{\varpi}\ge 0, \boldsymbol{\xi}\ge 0,\\
			&\hphantom {s.t.~}
			|{v}_m|^2 = 1, \forall m,	\label{robust1_RIS1_amplitude}
		\end{alignat} 
	\end{subequations} 
	where the Modified-(\ref{robust1_beamforming1_Com11_lower3}) constraints are LMIs revised from~\eqref{robust1_beamforming1_Com11_lower3} by replacing $z_k\gamma_k$ with $z_k\eta$, which leads to the non-convexity of the problem in~\eqref{robust1_RIS1}. The following lemma could be employed to tackle this challenge. 
	
	\textit{Lemma 2:} For any complex hemitian matrix $\boldsymbol{A} \in \mathbb{C}^{N\times N},\boldsymbol{b}\in\mathbb{C}^{N\times1},c\in\mathbb{C}$, and $d\in\mathbb{C}$, it follows that 
	\begin{equation}
		\begin{bmatrix}
			\boldsymbol{A} & \boldsymbol{b} \\ \boldsymbol{b}^H & c
		\end{bmatrix}
		\succeq \boldsymbol{0}
		\Rightarrow
		\begin{bmatrix}
			\boldsymbol{A} & \boldsymbol{b} \\ \boldsymbol{b}^H & d
		\end{bmatrix}
		\succeq \boldsymbol{0},
	\end{equation}
	holds if and only if $d\geq c$. 
	
	Bases on \textit{Lemma 2}, the Modified-(\ref{robust1_beamforming1_Com11_lower3}) constraints can be equivalently recast as 
	\begin{align} 
		&\begin{bmatrix} 
			\varpi _{k}\boldsymbol {I}_{M}+\boldsymbol{X}_{k} & \boldsymbol{X}_k^H\hat{\boldsymbol{h}}_k \\ 
			\hat{\boldsymbol{h}}_k^H\boldsymbol{X}_k &c_k
		\end{bmatrix} \succeq \mathbf {0},\forall k,\label{robust1_RIS_S_lemma}\\
		& \hat{\boldsymbol{h}}_k^H\boldsymbol{X}_k\hat{\boldsymbol{h}}_k - z_k\eta-\varpi _{k}\varepsilon_k^2 \ge c_k, \forall k,\label{robust1_RIS_c}
	\end{align} 
	where $\boldsymbol{c} = [c_1,\dots,c_K]^T\in\mathbb{C}^{K\times1}$ is the auxiliary variable. It should be noted that the second term in the LHS of the constraints in~\eqref{robust1_RIS_c}, i.e., $z_k\eta$, is not jointly convex with respect to (w.r.t) $\eta$ and $z_k$~\cite{7547360}, but it satisfies
	\begin{equation}
		\eta z_k\le \frac{1}{2}\left(\frac{z_k^{(l)}}{\eta^{(l)}}\eta^2+\frac{\eta^{(l)}}{z_k^{(l)}}z_k^2\right), \forall k.\label{etaz}
	\end{equation}
	Thus, the constraints in~\eqref{robust1_RIS_c}	can be approximated as
	\begin{equation}
		\hat{\boldsymbol{h}}_k^H\boldsymbol{X}_k\hat{\boldsymbol{h}}_k - \frac{1}{2}\left(\frac{z_k^{(l)}}{\eta^{(l)}}\eta^2+\frac{\eta^{(l)}}{z_k^{(l)}}z_k^2\right)-\varpi _{k}\varepsilon_k^2 \ge c_k, \forall k.\label{zk_approximate}
	\end{equation}
	
	Then, we move on to deal with the unit-modulus constraints in~\eqref{robust1_RIS1_amplitude} by using the penalty technique. Specifically, we revise the problem in~\eqref{robust1_RIS1} to be a penalized version as 
	\begin{subequations}\label{robust1_RIS2}
		\begin{alignat}{2}
			&\underset{ \boldsymbol{V},\boldsymbol{\varpi},\boldsymbol{\xi},\boldsymbol{z},\boldsymbol{c},\eta}{\max} ~ \eta + \rho_1(||\boldsymbol{v}||^2-M)\label{robust1_RIS2_OB}\\
			&\,\text {s.t.}~\text{(\ref{robust1_beamforming1_Com22_lower2}),~(\ref{robust1_RIS_S_lemma}),~(\ref{zk_approximate})},\boldsymbol{\varpi}\ge 0, \boldsymbol{\xi}\ge 0,\\
			&\hphantom {s.t.~}
			|{v}_m|^2 \le 1, \forall m,	\label{robust2_RIS1_amplitude}
		\end{alignat} 
	\end{subequations} 
	where $\boldsymbol{v} = [v_1,v_2,\dots,v_M]^T\in \mathbb{C}^{M\times1}$, and $\rho_1$ is a positive constant which encourages the solution of~\eqref{robust1_RIS2} to satisfy~\eqref{robust1_RIS1_amplitude}. Note that the penalty term $\rho_1 \left(||\boldsymbol{v}||^2 - M\right)$ in (\ref{robust1_RIS2_OB}) leads to a non-concave objective function, therefore we follow the principle of convex successive approximation method to iteratively approximate (\ref*{robust1_RIS2_OB}) with its first-order Taylor expansion. Consequently, the problem in (\ref{robust1_RIS2}) can be approximated as
	\begin{subequations}\label{robust_RIS_approximated2}
		\begin{alignat}{2}
			&\underset{\boldsymbol{V},\boldsymbol{\xi},\boldsymbol{\varpi},\boldsymbol{z},\boldsymbol{c},\eta}{\max} ~\eta + \rho_1 \Re\{2(\boldsymbol{v}^{(l)})^H\boldsymbol{v} -(\boldsymbol{v}^{(l)})^H\boldsymbol{v}^{(l)} \} - \rho_1M\\
			&\,\text {s.t.}~\text{(\ref{robust1_beamforming1_Com22_lower2}),~(\ref{robust1_RIS_S_lemma}),~(\ref{zk_approximate}),~(\ref{robust2_RIS1_amplitude})},\boldsymbol{\varpi}\ge 0, \boldsymbol{\xi}\ge 0,
		\end{alignat} 
	\end{subequations} 
	which is  an SDP and can be efficiently solved by the CVX tool~\cite{boyd2004convex}. A two-layer loop is  adopted for (\ref{robust_RIS}). In the outer loop, the penalty factor $\rho_1$ is initially set to a small value, e.g., $10^{-2}\sim10^{-1}$ times the RIS size $M$, to find a proper starting point, then updated until sufficiently large. In the inner loop, the problem in~\eqref{robust_RIS_approximated2} is solved iteratively to update $\boldsymbol{v}$ and $\eta$ with $\rho_1$ fixed. The steps are summarized in \textit{Algorithm 1}.
	\begin{algorithm}[t]
		\caption{Penalty Optimization on Reflecting Coefficients}
		\begin{algorithmic}[1] \STATE \textbf{Initialize:} set $l=0,\rho_1>0,\tau>1$, and initialize 
			$\boldsymbol{v}^{(l)}$.		
			\REPEAT	
			
			\STATE $\rho_1$ = $\tau \rho_1$.
			
			\REPEAT	
			
			\STATE Update $\boldsymbol{v}^{(l+1)}$ from the problem in (\ref{robust_RIS_approximated2}).	
			\STATE $l=l+1$.	
			\UNTIL $||\boldsymbol{v}^{(l)} - \boldsymbol{v}^{(l-1)}||^2 \le \xi_1$.
			
			\UNTIL $(||\boldsymbol{v}^{(l)}||^2 - M) \le \xi_2$.
			\RETURN  $\boldsymbol{v}^\star = \boldsymbol{v}^{(l)}$.
		\end{algorithmic} 
	\end{algorithm}
	\subsubsection{Updating Antenna Positions}
	
	In this subsection, we focus on the optimization on  antenna positions.
	We only need to show the robust design for positions of transmitting MAs $\tilde{\boldsymbol{t}}$, as the design for $\tilde{\boldsymbol{r}}$ is similar and thus omitted for brevity. 
	
	In order to expose $\tilde{\boldsymbol{t}}$ in (\ref{robust1_surrogate_reformulated}), 
	the objective function $\bar{\Gamma}_r(\tilde{\boldsymbol{t}})$ is first transformed into a more explicit form, as shown in (\ref{robust1_surrogate_reformulated1}) at the top of next page. 
	\begin{figure*}[tp]	 	\begin{align}\label{robust1_surrogate_reformulated1}
			\bar{\Gamma}_r(\tilde{\boldsymbol{t}}) = ~ 2\zeta_0^2\Re\{\sum_{p=1}^{P_s}\mu_p&\avector{p}{r}^H\boldsymbol{\Lambda}\boldsymbol{W}^H\avector{p}{t}\} - \notag \\ &\sum_{q=1}^{Q}\zeta_0^2\zeta_q^2\avector{q}{r}^H\boldsymbol{\Lambda}\boldsymbol{\Lambda}^H\avector{q}{r}\avector{q}{t}^H\boldsymbol{W}\boldsymbol{W}^H\avector{q}{t}.
		\end{align}
		\hrule
	\end{figure*}
	Note that the terms independent of $\tilde{\boldsymbol{t}}$ in~\eqref{robust1_surrogate_reformulated1} are omitted. Letting $\boldsymbol{b}_p =\mu_p\zeta_0^2 \boldsymbol{W}\boldsymbol{\Lambda}^H\avector{p}{r},c_q = \zeta_0^2\zeta_q^2\avector{q}{r}^H\boldsymbol{\Lambda}\boldsymbol{\Lambda}^H\avector{q}{r},$ and $\boldsymbol{D} = \boldsymbol{W}\boldsymbol{W}^H$, $\bar{\Gamma}_r(\tilde{\boldsymbol{t}})$ can be reformulated as 
	\begin{align}
		\bar{\Gamma}_r(\tilde{\boldsymbol{t}}) = 2\Re\{&\sum_{p=1}^{P_s}\boldsymbol{b}_p^H\avector{p}{t}\} -\notag\\ &\sum_{q=1}^{Q}c_q\avector{q}{t}^H\boldsymbol{D}\avector{q}{t}.
	\end{align}
	Thus, the problem in (\ref{robust_P1}) is recast as
	\begin{subequations}\label{robust3}
		\begin{alignat}{2}
			&\underset{\tilde{\boldsymbol{t}}}{\max}~~\bar{\Gamma}_r(\tilde{\boldsymbol{t}})\label{robust3_radar}\\
			&\,\text {s.t.}~ \Gamma_k(\tilde{\boldsymbol{t}}) \ge \gamma_k,||\Delta\boldsymbol{h}_k||_2\le \varepsilon_k,\forall k,\label{robust3_com}\\
			&\hphantom {s.t.~}||\boldsymbol{t}_n-\boldsymbol{t}_{n'}||_2^2 \ge D^2,\forall n \neq n',\label{robust3_MA1}\\
			&\hphantom{s.t.~}\boldsymbol{t}_n \in \mathcal{C}_t.\label{robust3_MA2}
		\end{alignat} 
	\end{subequations}
	The problem in (\ref{robust3}) is intractable due to the non-convexity of the objective function in (\ref{robust3_radar}), the constraints in  (\ref{robust3_com}), as well as  (\ref{robust3_MA1}). To deal with these issues, the SCA method can be applied\cite{razaviyayn2014successive}. We first deal with the non-convexity of  the objective function $\bar{\Gamma}_r(\tilde{\boldsymbol{t}})$ in~\eqref{robust3_radar}. Specifically,  according to the second-order Taylor expansion theorem in~\cite{razaviyayn2014successive}, $\bar{\Gamma}_r(\tilde{\boldsymbol{t}})$ can be lower bounded by a quadratic surrogate concave function, i.e.,
	\begin{equation}\label{robust3_radar_approximated}
		\bar{\Gamma}_r(\tilde{\boldsymbol{t}})\geq \bar{\Gamma}_r(\tilde{\boldsymbol{t}}^{(l)})+\nabla \bar{\Gamma}_r(\tilde{\boldsymbol{t}}^{(l)})^T(\tilde{\boldsymbol{t}}-\tilde{\boldsymbol{t}}^{\left(l\right)})-\frac{\delta_0}{2}(\tilde{\boldsymbol{t}}-\tilde{\boldsymbol{t}}^{\left(l\right)})^T(\tilde{\boldsymbol{t}}-\tilde{\boldsymbol{t}}^{\left(l\right)}),
	\end{equation}
	where $\tilde{\boldsymbol{t}}^{(l)}$ denotes the positions of MAs obtained in the $l$-th iteration, $\nabla \bar{\Gamma}_r(\tilde{\boldsymbol{t}}^{(l)}) \in \mathbb{C}^{2N \times 1}$   is the gradient vector at $\tilde{\boldsymbol{t}}^{(l)}$, and  $\delta_0$ is a positive real number satisfying $\delta_0 \boldsymbol{I}_{2N}\succeq \nabla^2\bar{\Gamma} _r(\tilde{\boldsymbol{t}}^{(l)})$ with the Hessian matrix $\nabla^2\bar{\Gamma} _r(\tilde{\boldsymbol{t}}^{(l)}) \in \mathbb{C}^{2N \times 2N}$. Please refer to Appendix A for the construction of $\nabla \bar{\Gamma}_r(\tilde{\boldsymbol{t}}^{(l)})$, $\delta_0$, and $\nabla^2\bar{\Gamma} _r(\tilde{\boldsymbol{t}}^{(l)})$.

	Next, recalling that $\boldsymbol{H}(\tilde{\boldsymbol{t}}) = \boldsymbol{F}({\tilde{\boldsymbol{s}}})^H\boldsymbol{\Sigma}\boldsymbol{G}(\tilde{\boldsymbol{t}})$,  the constraints in (\ref{robust3_com}) can be first recast as
	\begin{equation}
		\underbrace{\boldsymbol{a}_k^H \boldsymbol{G}(\tilde{\boldsymbol{t}})\boldsymbol{R}_k\boldsymbol{G}(\tilde{\boldsymbol{t}})^H \boldsymbol{a}_k}_{\triangleq {f}_k(\tilde{\boldsymbol{t}})} + \gamma_k \sigma_k^2 \leq 0, ||\Delta \boldsymbol{h}_k||\le \varepsilon_k, \forall k,
	\end{equation}
	where  $\boldsymbol{R}_k = \sum_{j\neq k}^{K}\gamma_k\boldsymbol{w}_j\boldsymbol{w}_j^H - \boldsymbol{w}_k\boldsymbol{w}_k^H\in\mathbb{C}^{N\times N}$ and $\boldsymbol{a}_k = \boldsymbol{\Sigma}^H\boldsymbol{F}(\boldsymbol{s})\boldsymbol{V}^H\boldsymbol{h}_k\in\mathbb{C}^{L\times1}$. As $f_k(\tilde{\boldsymbol{t}})$ is neither convex nor concave w.r.t. $\tilde{\boldsymbol{t}}$, we construct a surrogate function that serves as an upper bound of $f_k(\tilde{\boldsymbol{t}})$ based on the second-order Taylor expansion~\cite{razaviyayn2014successive}, i.e., 
	\begin{equation}
		f_k(\tilde{\boldsymbol{t}}) \le 	f_k(\tilde{\boldsymbol{t}}^{\left(l\right)}) + \nabla 	f_k(\tilde{\boldsymbol{t}}^{\left(l\right)})^T (\tilde{\boldsymbol{t}} - \tilde{\boldsymbol{t}}^{\left(l\right)}) +\frac{\delta_k}{2}(\tilde{\boldsymbol{t}} - \tilde{\boldsymbol{t}}^{\left(l\right)})^T(\tilde{\boldsymbol{t}} - \tilde{\boldsymbol{t}}^{\left(l\right)}),\label{upper}
	\end{equation}
	where $\nabla f_k(\tilde{\boldsymbol{t}}) \in \mathbb{C}^{2N\times 1}$ denotes the gradient vector over $\tilde{\boldsymbol{t}}$. A positive real number $\delta_k$ is selected to satisfy $\delta_k \boldsymbol{I}_{2N}\succeq \nabla^2 	f_k(\tilde{\boldsymbol{t}})$ with  $\nabla^2 f_k(\tilde{\boldsymbol{t}})\in \mathbb{C}^{2N\times2N}$ being the  Hessian matrix. Please refer to Appendix B for the construction of $\nabla f_k(\tilde{\boldsymbol{t}}), \delta_k$, and $\nabla^2 f_k(\tilde{\boldsymbol{t}})$. 
	
	With the aid of  $\boldsymbol{h}_k = \hat{\boldsymbol{h}}_k + \Delta\boldsymbol{h}_k$, the upper bound  of $f_k(\tilde{\boldsymbol{t}})$ in~\eqref{upper}  can be reformulated as
	\begin{align}\label{transformation}
		f_k&(\tilde{\boldsymbol{t}}) \le f_k(\tilde{\boldsymbol{t}}^{\left(l\right)}) + \boldsymbol{h}_k^H\boldsymbol{\Phi}_k\boldsymbol{h}_k +\frac{\delta_k}{2}(\tilde{\boldsymbol{t}} - \tilde{\boldsymbol{t}}^{\left(l\right)})^T(\tilde{\boldsymbol{t}} - \tilde{\boldsymbol{t}}^{\left(l\right)})\notag\\
		& = \Delta\boldsymbol{h}_k^H(\boldsymbol{T}_k+\boldsymbol{\Phi}_k)\Delta\boldsymbol{h}_k^H+2\Re\{\hat{\boldsymbol{h}}_k^H(\boldsymbol{T}_k+\boldsymbol{\Phi}_k)\Delta\boldsymbol{h}_k\}\notag\\
		&+\hat{\boldsymbol{h}}_k^H(\boldsymbol{T}_k+\boldsymbol{\Phi}_k)\hat{\boldsymbol{h}}_k +\frac{\delta_k}{2}(\tilde{\boldsymbol{t}} - \tilde{\boldsymbol{t}}^{\left(l\right)})^T(\tilde{\boldsymbol{t}} - \tilde{\boldsymbol{t}}^{\left(l\right)}),\forall k,
	\end{align}
	where $\boldsymbol{T}_k = \boldsymbol{V}\boldsymbol{H}(\tilde{\boldsymbol{t}}^{(l)})\boldsymbol{R}_k\boldsymbol{H}(\tilde{\boldsymbol{t}}^{(l)})^H\boldsymbol{V}^H$, and the transformation in (\ref{transformation})  and the derivations of $\{\boldsymbol{\Phi}_k\}_{k=1}^K$   can be found in Appendix C. Based on the S-Lemma, the above inequality in~(\ref{transformation}) can be further given by
	\begin{align}
		\left [{\begin{array}{cc} \varpi _{k}\boldsymbol{I}_M+\boldsymbol{T}_k+\boldsymbol{\Phi}_k &~~ (\hat{\boldsymbol{h}}_k^H\boldsymbol{T}_k+\hat{\boldsymbol{h}}_k^H\boldsymbol{\Phi}_k)^H\\ \hat{\boldsymbol{h}}_k^H\boldsymbol{T}_k+\hat{\boldsymbol{h}}_k^H\boldsymbol{\Phi}_k &~~ d_k - \gamma_k\sigma_k^2 -\varpi_k\varepsilon_k^2 \end{array}}\right]\succeq \boldsymbol{0}_{M+1},\forall k,\label{robust3_com_lower1}
	\end{align}
	where $d_k = \hat{\boldsymbol{h}}_k^H(\boldsymbol{T}_k+\boldsymbol{\Phi}_k)\hat{\boldsymbol{h}}_k + \frac{\delta_k}{2}(\tilde{\boldsymbol{t}} - \tilde{\boldsymbol{t}}^{\left(l\right)})^T(\tilde{\boldsymbol{t}} - \tilde{\boldsymbol{t}}^{\left(l\right)})$,  and $\varpi _{k}\ge0$ is the auxiliary variable. However,  the quadratic term in $d_k$, i.e.,  $  \frac{\delta_k}{2}(\tilde{\boldsymbol{t}} - \tilde{\boldsymbol{t}}^{\left(l\right)})^T(\tilde{\boldsymbol{t}} - \tilde{\boldsymbol{t}}^{\left(l\right)})$, leads to the non-convexity of~(\ref{robust3_com_lower1}). By introducing an auxiliary variable $\boldsymbol{\iota} = [\iota_1,\dots,\iota_K]^T$, we employ \textit{Lemma 2} to transform (\ref{robust3_com_lower1}) into the following LMIs
	\begin{align}
		\left [{\begin{array}{cc} \varpi _{k}\boldsymbol{I}_M+\boldsymbol{T}_k+\boldsymbol{\Phi}_k &~~ (\hat{\boldsymbol{h}}_k^H\boldsymbol{T}_k+\hat{\boldsymbol{h}}_k^H\boldsymbol{\Phi}_k)^H\\ \hat{\boldsymbol{h}}_k^H\boldsymbol{T}_k+\hat{\boldsymbol{h}}_k^H\boldsymbol{\Phi}_k &~~ \iota_k \end{array}}\right]\succeq \mathbf {0}_{M+1},\forall k,\label{robust3_com_lower2}
	\end{align}
	where  $\iota_k \le d_k - \gamma_k\sigma_k^2 -\varpi_k\varepsilon_k^2,\forall k$. 
	
	For the constraints in~\eqref{robust3_MA1}, since the term $\left\| \boldsymbol{t}_n - \boldsymbol{t}_{n'} \right\|_2^2$ is a  convex function w.r.t. $\boldsymbol{t}_n - \boldsymbol{t}_{n'}$, it can lower bounded by its first-order Taylor expansion at the given points $\boldsymbol{t}_n^{\left(l\right)}$ and $\boldsymbol{t}_{n'}^{\left(l\right)}$, i.e, 
	\begin{equation}
		\left\| \boldsymbol{t}_n - \boldsymbol{t}_{n'} \right\|_2^2 \geq 
		-\left\| \boldsymbol{t}_n^{(l)} - \boldsymbol{t}_{n'}^{(l)} \right\|_2^2 + 
		2( \boldsymbol{t}_n^{(l)} - \boldsymbol{t}_{n'}^{(l)})^T  
		( \boldsymbol{t}_n - \boldsymbol{t}_{n'}), \label{robust3_MA1_APP}
	\end{equation}
	where $1\le n\neq n'\le N$. Thus, combing~\eqref{robust3_radar_approximated}, \eqref{robust3_com_lower2}, and \eqref{robust3_MA1_APP}, we obtain the approximated optimization problem as follows
	\begin{small}
			\begin{subequations}\label{robust4}
			\begin{alignat}{2}
				&\underset{\tilde{\boldsymbol{t}},\boldsymbol{\varpi},\boldsymbol{\iota}}{\max}~ \bar{\Gamma}_r(\tilde{\boldsymbol{t}}^{(l)})+\nabla \bar{\Gamma}_r(\tilde{\boldsymbol{t}}^{(l)})^T(\tilde{\boldsymbol{t}}-\tilde{\boldsymbol{t}}^{\left(l\right)})-\frac{\delta_0}{2}(\tilde{\boldsymbol{t}}-\tilde{\boldsymbol{t}}^{\left(l\right)})^T(\tilde{\boldsymbol{t}}-\tilde{\boldsymbol{t}}^{\left(l\right)})\\
				&\,\text {s.t.}~\iota_k \le d_k - \gamma_k\sigma_k^2 -\varpi_k\varepsilon_k^2,\varpi_k\ge 0,\forall k.\\
				&\hphantom{s.t.~}~-\left\| \boldsymbol{t}_n^{(l)} - \boldsymbol{t}_{n'}^{(l)} \right\|_2^2 + 
				2( \boldsymbol{t}_n^{(l)} - \boldsymbol{t}_{n'}^{(l)})^T  
				( \boldsymbol{t}_n - \boldsymbol{t}_{n'})\ge D^2,\forall n\neq n',\\
				&\hphantom{s.t.~}~\text{(\ref{robust3_MA2}), (\ref{robust3_com_lower2}),}
			\end{alignat}
		\end{subequations}
	\end{small}
	which is an SDP and can be efficiently solved by the interior point method (IPM)~\cite{boyd2004convex}.
	\subsection{Optimization on (P2)}
	Letting $\boldsymbol{Z}_p =\zeta_0\boldsymbol{W}^H\boldsymbol{A}_p(\tilde{\boldsymbol{r}},\tilde{\boldsymbol{t}})^H (\boldsymbol{\Xi}+\sigma_r^2\boldsymbol{I}_N)^{-\frac{1}{2}} \in \mathbb{C}^{K\times N}$, the problem in (P2)
	can be rewritten as 
	\begin{subequations}\label{robust_mu}
		\begin{alignat}{2}
			& ~\underset{\boldsymbol{\mu}}{\min}  ~~\tilde{\Gamma} _r(\boldsymbol{\mu})=||\sum_{p=1}^{P_s}\mu_p\boldsymbol{Z}_p||^2_F,\\
			&\,\text {s.t.}~ \sum_{\mu=1}^{P_s}\mu_k=1,\mu_k\ge 0,
		\end{alignat}
	\end{subequations} 
	which is convex and can be solved by the CVX tool \cite{6891348}.
	\subsection{Convergence and Complexity Analysis}
	The two-layer BCD-based algorithm for solving the problem in~(\ref{robust1}) is summarized in \textit{Algorithm~2}. We outline that repeating procedures in the inner loop, i.e., the Steps 3-10, yields a non-decreasing sequence of the objective values of the problem in~\eqref{robust_P1} as the iterative procedures belong to the majorization-minimization (MM) method~\cite{7547360}. As a result, the inner loop is assured to converge at a suboptimal point.
	For the outer loop, each subproblem in \textit{Algorithm~2} is solved locally or optimally.  According to \cite{razaviyayn2020nonconvex}, \textit{Algorithm~2} is guaranteed to converge to a stationary point for the problem in~(\ref{robust1}). The computational complexity of each optimization variable is summarized in~Table \ref{table1}, shown at the top of this page. 
	\begin{table*}[t]
		\centering
		\caption{Complexity of the proposed algorithm.}\label{table1}
		\renewcommand{\arraystretch}{2.5}
		\begin{tabular}{c|c}
			\hline
			Optimization Variable  & Computational Complexity \\
			\hline
			$\boldsymbol{W}$ & $\mathcal{O}(\ln(\frac{1}{\epsilon_0})\sqrt{K^2+2KM}(N^2K^5+2MN^2K^4+M^2N^2K^3+N^3K^3+M^3NK^2))$  \\
			\hline
			$\boldsymbol{V}$ & $\mathcal{O}(\ln(\frac{1}{\epsilon_0})\sqrt{K^2+2KM}(KM^4+K^2M^3+K^3M^2+K^4M+K^5))$  \\
			\hline
			$\tilde{\boldsymbol{t}}$ & $\mathcal{O}(\ln(\frac{1}{\epsilon_0})\sqrt{N^2+KM}(N^4+KN^3+K^2N^2+K(N+K)^2M^2+K(N+K)M^3))$  \\
			\hline
			$\tilde{\boldsymbol{r}}$ & $\mathcal{O}(\ln(\frac{1}{\epsilon_0})N^4)$  \\
			\hline
			${\boldsymbol{\mu}}$ & $\mathcal{O}(\ln(\frac{1}{\epsilon_0})P_s^{3.5})$  \\
			\hline
			$\boldsymbol{\Lambda}$ & $\mathcal{O}(N^3)$  \\
			\hline
		\end{tabular}
	\end{table*}
	\begin{algorithm}[t] 
		\caption{Two-layer BCD Optimization Algorithm for (\ref{robust1})}
		\label{alg2}
		\begin{algorithmic}[1]
			\STATE  \textbf{Initialize}  $\boldsymbol{W}^{[\psi_1]},\boldsymbol{V}^{[\psi_1]},\tilde{\boldsymbol{t}}^{[\psi_1]},\tilde{\boldsymbol{r}}^{[\psi_1]},\boldsymbol{\Lambda}^{[\psi_1]},\boldsymbol{\mu}^{[\psi_2]},$  and set the inner-layer and outer-layer iteration indices $\psi_1 = \psi_2 = 0$.
			\REPEAT
			\REPEAT
			\STATE Update $\boldsymbol{\Lambda}^{[\psi_1+1]}$  via  (\ref{robust_Lambda});
			\STATE Update $\boldsymbol{W}^{[\psi_1+1]}$  by  solving (\ref{robust1_beamforming2});
			\STATE Update $\boldsymbol{V}^{[\psi_1+1]}$ by solving (\ref{robust_RIS}) via \textit{Algorithm 1};
			\STATE Update $\tilde{\boldsymbol{t}}^{[\psi_1+1]}$ by solving (\ref{robust4});
			\STATE Update $\tilde{\boldsymbol{r}}^{[\psi_1+1]}$ in a similar fashion as  $\tilde{\boldsymbol{t}}^{[\psi_1+1]}$;
			\STATE Let $\psi_1 = \psi_1 + 1$;
			\UNTIL The objective value (\ref{robust1_surrogate_reformulated}) converges.
			\STATE Update $\boldsymbol{\mu}$  by solving (\ref{robust_mu});
			\STATE Set $\psi_2 = \psi_2 + 1$ and $\psi_1 = 0$;
			\UNTIL The objective value (\ref{robust1_objective}) converges.
			\STATE Return $\boldsymbol{\Lambda}^\star,\boldsymbol{V}^\star,\boldsymbol{W}^\star,\tilde{\boldsymbol{t}}^\star,\tilde{\boldsymbol{r}}^\star$, and $\boldsymbol{\mu}^\star$.
		\end{algorithmic}
	\end{algorithm}
	\subsection{Performance Upper Bound} 
	In order to assess the robust dual-task performance of the proposed scheme, we further consider the joint transceiver design with perfect CSI to serve as  a  performance upper bound. Specifically, letting $(\varphi_0^e,\varphi_0^a)$ denote the accurate location of the sensing target and $\varepsilon_k = 0$, the optimization problem in~\eqref{robust1} can be simplified as 
	\begin{subequations}\label{Problem1}
		\begin{alignat}{2}
			&\underset{ \boldsymbol{W},\boldsymbol{V}, \tilde{\boldsymbol{r}},\tilde{\boldsymbol{t}}}{\max} \quad \Gamma_r(\boldsymbol{W},\tilde{\boldsymbol{r}},\tilde{\boldsymbol{t}})\label{P1_objective}\\
			&\,\text {s.t.}~ \Gamma_k(\boldsymbol{W},\boldsymbol{V},\tilde{\boldsymbol{t}}) \ge \gamma_k,\forall k,\label{P1_Com}\\
			&\hphantom {s.t.~}\text{(\ref{robust1_MA1}),~(\ref{robust1_MA2}),~(\ref{robust1_Pt}),~(\ref{robust1_RIS})},
		\end{alignat} 
	\end{subequations} 
	the solution to which has been  elaborated in our preliminary work in~\cite{yang2025joint}, and thus omitted for brevity. To  improve the proposed transceiver design framework, we further present a complexity analysis of the optimization algorithm in~\cite{yang2025joint}. Specifically, the computational complexity for updating $\boldsymbol{\Lambda}$  is given by $\mathcal{O}(N^3)$.  The computational cost for  $\boldsymbol{W},\boldsymbol{V},\tilde{\boldsymbol{t}}$ and $\tilde{\boldsymbol{r}}$ is given by $\mathcal{O}\left(\ln(\frac{1}{\epsilon_0})N^3K^{4.5}\right)$, $\mathcal{O}\left(\ln(\frac{1}{\epsilon_0})\sqrt{2K+M}(KM^3+K^2M^2+K^3)\right)$, $\mathcal{O}\left(\ln(\frac{1}{\epsilon_0})\sqrt{N^2+4K}(N^4+4N^3K)\right)$, and $\mathcal{O}\left(\ln(\frac{1}{\epsilon_0})N^4\right)$, respectively, with $\epsilon_0$ being the prescribed accuracy. The algorithm complexity is reduced in comparison to that with imperfect CSI, which is due to the absence of channel errors.

	\section{Numerical Results}
	
	{In this section,  computer simulations are carried out to evaluate the performance of the proposed method.  We compare our scheme against several baseline schemes as outlined below.
		\begin{itemize}
			\item \textbf{Upper bound performance scheme:} With perfect CSI,  the optimization algorithm  in~\cite{yang2025joint} is employed to enhance the dual-task performance. 	The results with perfect CSI {serve} as a performance upper bound, and help to evaluate the corresponding robust designs.
			\item \textbf{Fixed position antenna (FPA)}: The BS for RIS-enhanced DFRC systems is equipped with  uniform planar arrays,  with $N$ fixed  transmitting/receiving antennas spaced with a distance of $\frac{\lambda}{2}$;
			\item \textbf{Random position antenna (RPA)}: The antennas at the BS are randomly distributed in the moving region under the constraint of the minimum distance $D$ between each other;
			\item \textbf{Random RIS}: The RIS phase shifts are generated randomly, following a uniform distribution within the range $[0,2\pi]$;
			\item \textbf{Greedy antenna selection (GAS)}: The moving regions are quantized into discrete ports spaced by $\frac{\lambda}{2}$. The greedy search algorithm is employed for the position optimization~\cite{10077503}\footnote{The computational complexity for $\tilde{\boldsymbol{t}}$ and $\tilde{\boldsymbol{r}}$ are respectively characterized by $\mathcal{O}(\ln(\frac{1}{\epsilon_0})NP_o\sqrt{K(M+1)}(K^2M^3+K^3M^2+4K^3))$ and $\mathcal{O}(\ln(\frac{1}{\epsilon_0})(P_sP_oN^2+QP_oN^3))$, with $P_o$ being the number of antenna ports.}.
			\item \textbf{Passive beamforming (PBF)}: This scheme assumes a fixed-position antenna array configuration and employs random transmit beamforming $\boldsymbol{W}$, while only the RIS reflection coefficients are optimized via \textit{Algorithm 1}.
	\end{itemize}}

	{In our simulation, we assume that the BS and the RIS are  located at (0, 0) m and (30, 5) m, respectively. The users are randomly distributed in  a circle centered at (30, 0) m with a radius of 3 m. The geometry channel model is employed for the communication links \cite{zhu2023modeling}, where the numbers of transmitting and receiving paths are identical, i.e., $L^t_k = L^r_k = L = 8, \forall k$. Under this condition, the PRM for each  link  is diagonal, i.e., $\boldsymbol{\Sigma}_k = \mathrm{diag}\{\sigma_{k,1},\dots,\sigma_{k,L}\}$ with  $\sigma_{k,l} \sim \mathcal{CN}\left(0,\frac{c_k^2}{L}\right)$. Note that $c_k^2 = C_0d_k^{-\alpha}$ denotes the large-scale path loss, where $C_0 = -30~ \text{dB}$, and the  path-loss exponents $\alpha$ for  the BS-RIS link, RIS-user link, and BS-target link are  given by 2.4, 2.8, and 2.6, respectively. The uncertain angle intervals $\Psi_e$ and $\Psi_a$ are uniformly sampled each 1 degree~\cite{su2022secure}. We further assume that $\Delta \vartheta_e = \Delta\vartheta_a = \Delta \vartheta$. The CSI error bounds are defined as $\varepsilon_k \triangleq \tau||\hat{\boldsymbol{h}}_k||_2,\forall k$, where $\tau\in[0,1)$ accounts for the relative amount of the CSI uncertainties~\cite{9180053}. The distance between the BS and the target is 20 m. Other parameters unless otherwise specified: $K = 3, N = 4, M = 16, P_t = 15~\text{dBW}, Q = 2 ,\vartheta_e = 30^{\circ}, \vartheta_a = 45^{\circ}, \Delta \vartheta = 4^{\circ},\sigma_k^2 = \sigma_r^2 = -80~\text{dBm},\gamma_k = \gamma = 10~\text{dB}, \tau = 0.02, \lambda = 0.1~\mathrm{m}, D = \frac{\lambda}{2}, A = 2\lambda,  \mathcal{C}_t = \mathcal{C}_r=[-\frac{A}{2},\frac{A}{2}]\times[-\frac{A}{2},\frac{A}{2}].$ The AoDs and AoAs for the clutters are given by $\{\psi_1^e = 120^{\circ}, \psi_1^a = 90^{\circ}\}$ and $\{\psi_2^e = 135^{\circ}, \psi_2^a = 60^{\circ}\}$, respectively. The initial spacing between adjacent antenna elements is set to be $0.8\lambda$. The simulation is performed on a computer with Core i9-10900K CPU and 32G RAM.}
	\begin{figure}[t]
		\centering
		\includegraphics[width=0.45\textwidth]{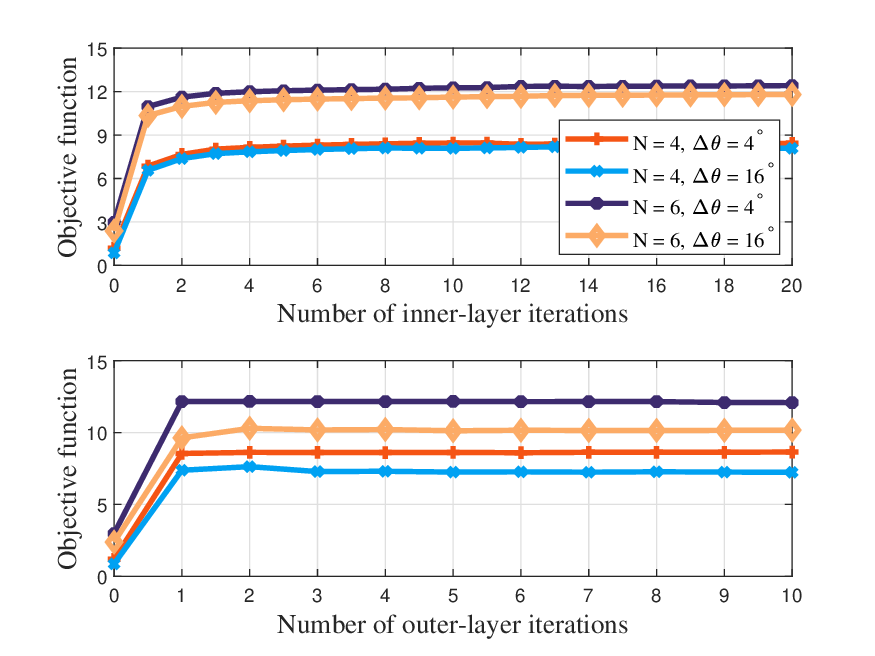}
		\captionsetup{font={normalsize},labelsep=period,singlelinecheck=off}
		\caption{Convergence behaviour of the two-layer BCD method.} 
		\label{convergence} 
	\end{figure}%
	\subsection{Convergence Behaviour and Algorithm Scalability}
	In Fig.\,\ref{convergence}, we present the convergence performance of \textit{Algorithm 2}. For the inner loop,  it can be seen that the minimum radar SINR for different angle errors and antenna numbers is monotonically increasing with the number of iterations, which is consistent with the behaviour of the MM method. In most cases, less than 6 iterations are sufficient. For the outer loop, the algorithm converges in approximately 4 iterations, which confirms the rapid convergence and superiority of the proposed algorithm. 
	
	{The scalability of the proposed algorithm is evaluated  by its running time under different system configurations, as summarized  in Table \ref{table2}. It should be noted that the number of RIS elements $M$ is the dominant factor influencing the computational overhead, despite the presence of higher-order terms of $K$ such as $\mathcal{O}(K^6)$ in Table \ref{table1}. For instance,  doubling $M$ from 32 to 64 with $K=4$ results in an 8.8 fold increase in running time from 50.4 s to 444.6 s, whereas increasing $K$ from 2 to 4 leads to only a moderate rise in execution time. This phenomenon can be attributed to two main reasons. Firstly, there is a significant scale disparity between the parameters. Specifically, the number of RIS elements $M$ is usually much larger than $K$ to guarantee passive beamforming gains. Consequently, the absolute value of $M^{4.5}$ significantly outweighs $K^6$, making the RIS-related computations the primary bottleneck. Secondly, the size of RIS $M$ directly determines the dimensionality of SDP problems, whereas $K$ primarily scales the number of QoS constraints. Despite the increased computational cost for large-scale RIS, the proposed algorithm maintains a reasonable execution time for moderate system configurations, verifying its feasibility for practical deployment.}

				\begin{table}
					\centering
					\renewcommand{\arraystretch}{1.7}
					\caption{Average running time (seconds) for different numbers of users and RIS elements.}
					\label{table2}
				\begin{tabular}{|c|c|c|c|c|c|}
					\hline
					\diagbox{User}{RIS} & $M$=4 & $M$=8 & $M$=16 & $M$=32 & $M$=64 \\
					\hline
					$K$=2 & 6.327 & 8.861
					& 15.495&32.394&159.455\\
					\hline
					$K$=4 & 7.769
					& 10.109
					& 24.320 &50.424&444.645\\
					\hline
				\end{tabular}
			\end{table}

	\subsection{Impact of Transmission Power}
	\begin{figure}[t]
		\centering
		\includegraphics[width=0.45\textwidth]{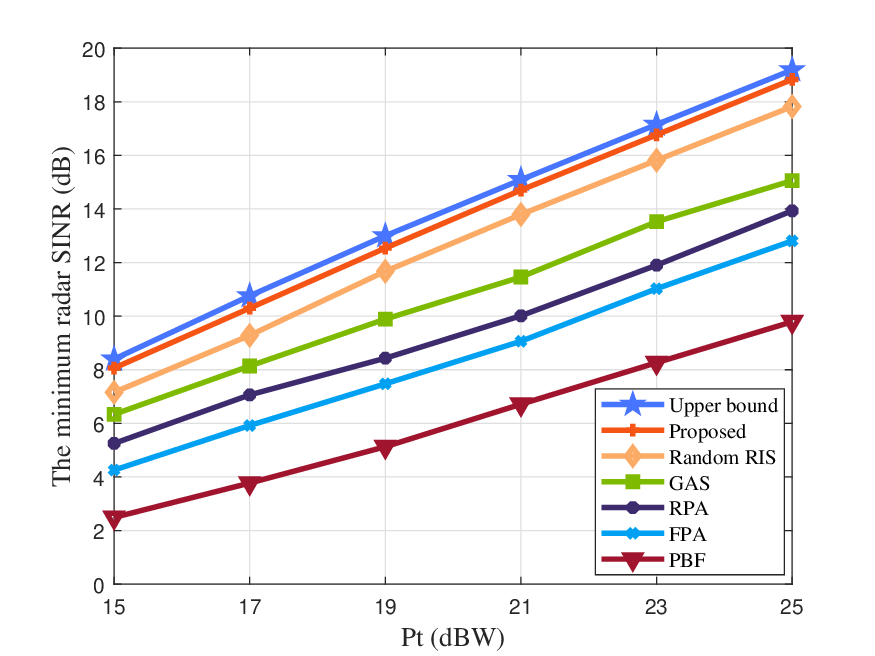}
		\captionsetup{font={normalsize},labelsep=period,singlelinecheck=off}
		\caption{The minimum radar SINR of different schemes versus  transmission power $P_t$.}
		\label{Pt} 
	\end{figure}%
	\begin{figure}[t]
		\centering
		\includegraphics[width=0.45\textwidth]{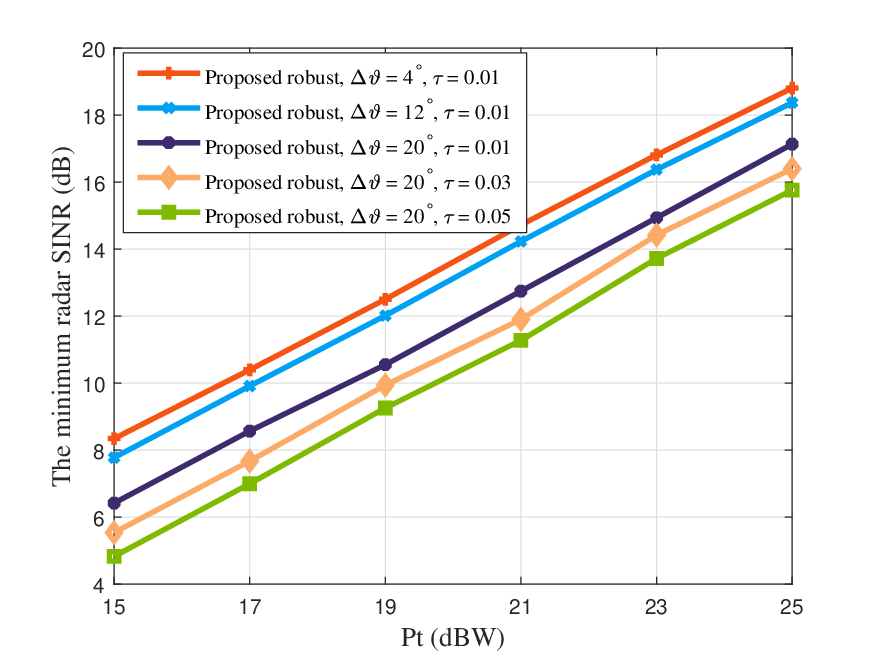}
		\captionsetup{font={normalsize},labelsep=period,singlelinecheck=off}
		\caption{The minimum radar SINR of different channel uncertainty levels versus  transmission power $P_t$.} 
		\label{robust_Pt} 
	\end{figure}%
	
	{Fig.\,\ref{Pt} shows the minimum radar SINR performance of different schemes versus the transmission power $P_t$. It can be seen that the proposed design only suffers about 0.5 dB  performance loss compared with the ideal case, which confirms the effectiveness and feasibility of the proposed robust solution. It is worth noting that the proposed design significantly 
		outperforms all other robust benchmark schemes in the sensing performance. This superiority can be attributed to  the active movement of antenna elements, which not only enhances the channel gain but also reconstructs the array geometry. In addition, the performance gap between the proposed scheme and the FPA scheme increases from 3.8 dB to 6 dB with increasing $P_t$. This is because the spatial DoFs offered by MAs enable a more flexible and effective power allocation between the antenna elements. Moreover, the Random RIS scheme experiences only about 1 dB SINR performance loss compared to the proposed scheme. This is reasonable as the radar SINR is independent of RIS, and the optimization on RIS is performed only to enhance the QoS performance. Furthermore, the PBF scheme with random transmit beamforming  yields the lowest minimum radar SINR among all considered schemes, further confirming the independence of the radar SINR from RIS configuration.
		In the GAS scheme, there is a limited number of available antenna ports in the moving region, which reduces the feasible region for position optimization. Besides, the greedy search algorithm generally only finds a good feasible solution, resulting in the performance degradation compared with the proposed scheme. }
	
	In Fig.\,\ref{robust_Pt}, we plot the minimum radar SINR of the proposed scheme with various channel uncertainty levels against the transmission power $P_t$.  
	We note that even under conditions where $\Delta\theta = 20^\circ$ and $\tau = 0.05$, the proposed scheme still maintains relatively satisfactory sensing performance, which indicates that our robust method can find applications even in DFRC scenarios with significant channel errors.
	\subsection{Impact of Moving Region Size}
	\begin{figure}[t]
		\centering
		\includegraphics[width=0.45\textwidth]{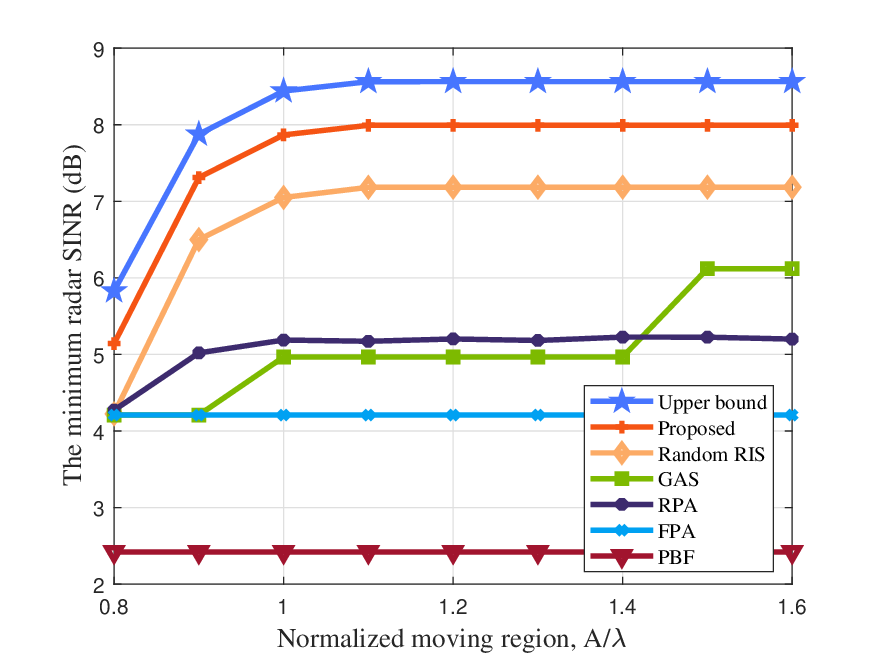}
		\captionsetup{font={normalsize},labelsep=period,singlelinecheck=off}
		\caption{The minimum radar SINR versus the normalized moving region.} 
		\label{region} 
	\end{figure}%
	{We then plot the minimum radar SINR versus the normalized region size $A/\lambda$ in Fig. \ref{region}. It is depicted that there is a significant gain in the minimum radar SINR of the proposed scheme as the  region size enlarges, surpassing the performance of other robust baseline designs. This is attributed to the enhanced flexibility in designing the antenna positions within a larger moving region. In particular,  we note that the performance enhancement brought by the enlarged region size is limited.  This result shows that the maximum  radar SINR is practically attainable within finite regions. Consequently, we argue that a moderately sized moving region can be selected to strike a satisfactory balance between the dual-task performance and hardware costs.}
	\subsection{Impact of QoS $\gamma$}
	\begin{figure}[t]
		\centering
		\includegraphics[width=0.45\textwidth]{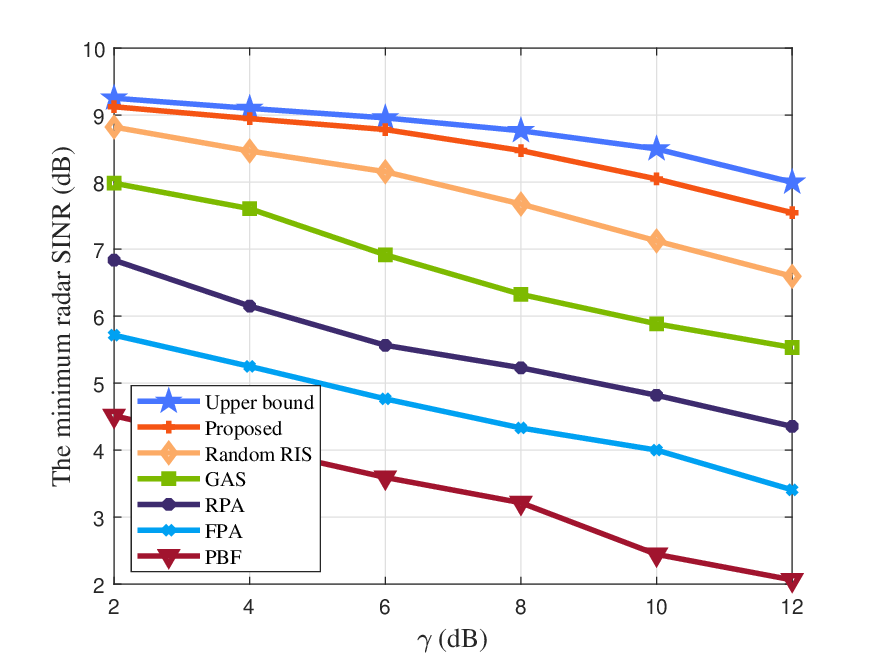}
		\captionsetup{font={normalsize},labelsep=period,singlelinecheck=off}
		\caption{The minimum radar SINR versus QoS $\gamma$.}
		\label{QoS} 
	\end{figure}%
	
	{We then evaluate the trade-off between the minimum radar SINR and communication QoS $\gamma$ in Fig. \ref{QoS}.  As the QoS $\gamma$ increases, the radar SINR of all schemes shows a declining trend. This is intuitive since a larger $\gamma$ requires more system resources to mitigate the inter-user interference. 
		It can be seen that the overall performance of the proposed method remains remarkably stable compared with other robust baseline schemes. This is because the proposed algorithm is able to leverage the diversity gain and the interference mitigation gain in the spatial domain by properly adjusting the antenna positions.
		In particular, the minimum radar SINR of the proposed scheme exhibits a minor decline from 9.1 dB to 7.5 dB as  $\gamma$ increases from 2 dB to 12 dB, while that of the Random RIS scheme has dropped from 8.8 dB to 6.6 dB. The performance gap enlarges because the random RIS coefficients fail to perform passive beamforming to enhance the  QoS performance. Consequently, more power must be allocated to meet the QoS requirements with increased communication loads, leading to  degradation of  radar performance. This observation highlights that the RIS is more effective in facilitating a better performance trade-off between radar and communications, rather than directly enhancing the radar SINR. Moreover, the minimum radar SINR of both the RPA and GAS schemes significantly deteriorates with increasing $\gamma$, which highlights the benefits of optimization on continuous antenna positions.
	}
	
	
	\subsection{Antenna Movement Analysis on Echo Signal Power Map}
	\begin{figure}[t]
		\centering
		\includegraphics[width=0.45\textwidth]{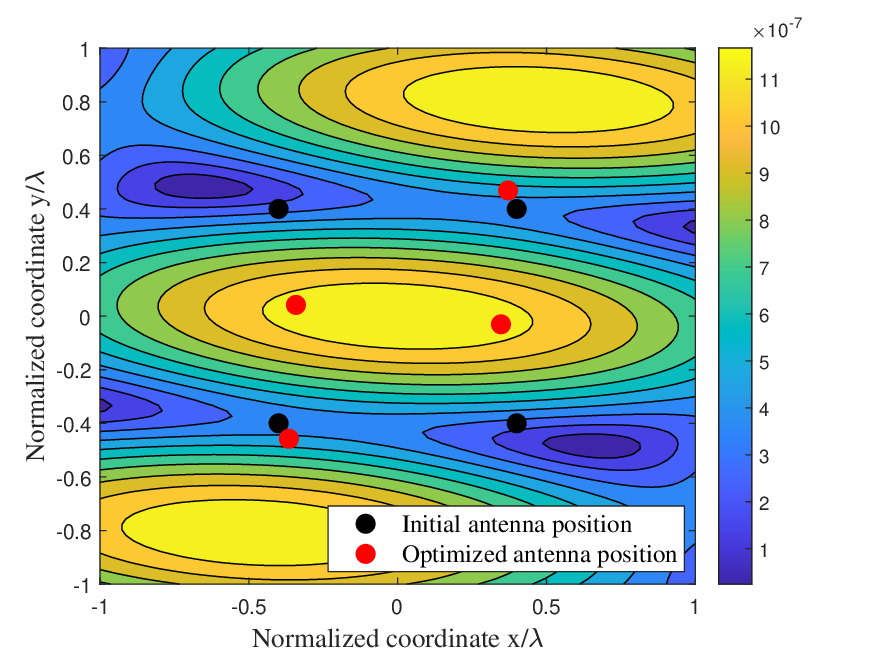}
		\captionsetup{font={normalsize},labelsep=period,singlelinecheck=off}
		\caption{Example of received signal power in Rx region $\mathcal{C}_r$.} 
		\label{energy} 
	\end{figure}%
	To characterize the impact of antenna movement on improving radar performance and reconstructing the array geometry, we further show the relation between the received echo signal power and positions of receiving MAs in Fig. \ref{energy}. To make it more intuitive, we assume that the target location is perfectly known by the BS. The results in Fig. \ref{energy} reveal significant position-dependent variations in echo signal power gains.
	Interestingly, we note that the  antenna elements are not all optimized to locations with the maximum signal power gain. This is because the optimization considers both enhancement of the target echo signal power  and suppression of clutter interference.  This observation demonstrates  that strategically optimizing antenna positions can significantly enhance the radar sensing performance.
	\subsection{Impact of Various Channel Uncertainty Levels}
	In this subsection, we evaluate the dual-task performance of different schemes under various channel uncertainty levels.
	\begin{figure}[t]
		\centering
		\includegraphics[width=0.45\textwidth]{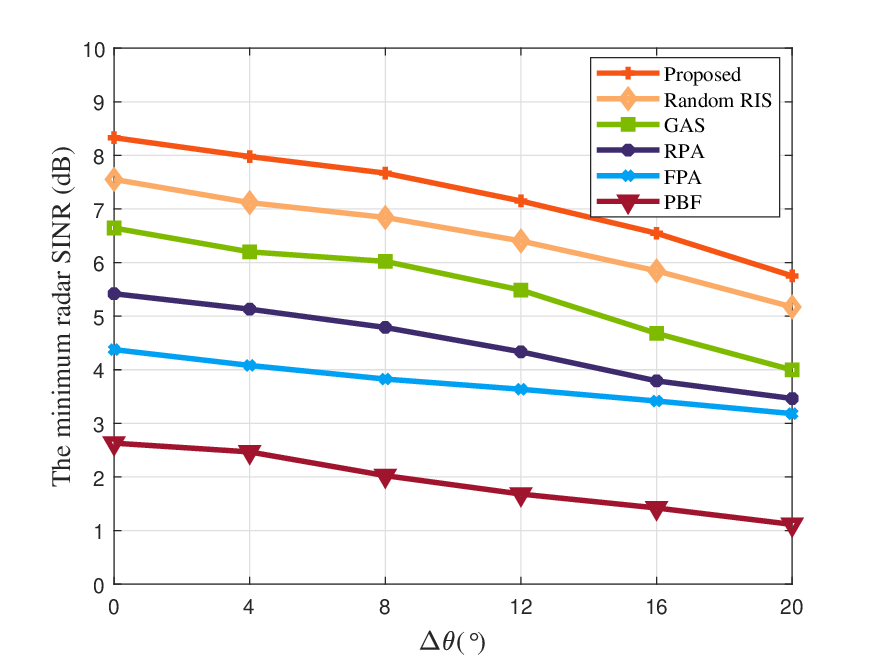}
		\captionsetup{font={normalsize},labelsep=period,singlelinecheck=off}
		\caption{The minimum radar SINR versus angle uncertainty $\Delta \vartheta$ with $\tau = 0.02$.} 
		\label{theta} 
	\end{figure}%
	
	{In Fig. \ref{theta}, we plot the minimum radar SINR versus the angle uncertainty $\Delta \vartheta$. It is evident that the proposed algorithm consistently achieves the highest minimum radar SINR across all levels of angle uncertainty. This is attributed to the spatial DoFs provided by MAs, which help maintain a higher SINR compared to the other methods. We observe that the benefits brought by MAs become less pronounced as $\Delta \vartheta$ increases. 
		This is because, with the increase of uncertain regions, it becomes more difficult to determine the target position, which reduces the gain brought by antenna movement. Nevertheless, it can be seen that the RPA scheme still outperforms the FPA scheme, which highlights the potential of movable antennas.}
	
	\begin{figure}[t]
		\centering
		\includegraphics[width=0.45\textwidth]{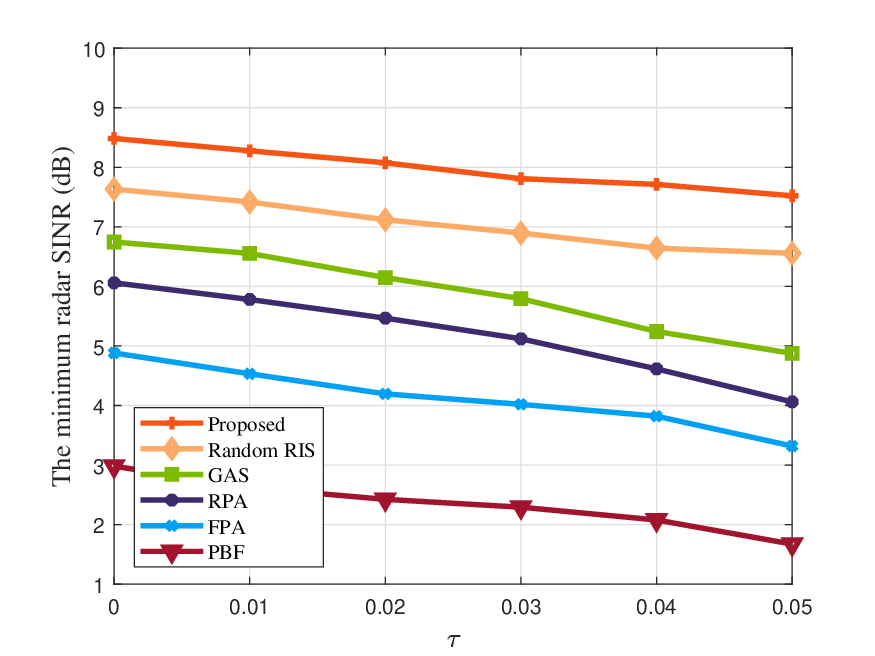}
		\captionsetup{font={normalsize},labelsep=period,singlelinecheck=off}
		\caption{The minimum radar SINR versus the communication channel uncertainty level $\tau$ with $\Delta \vartheta = 4^{\circ}$.}
		\label{delta} 
	\end{figure}%
	
{Fig.\,\ref{delta} illustrates the effect of communication channel errors  on the minimum radar SINR. For the bounded CSI error model, the optimization is performed to meet the given QoS requirements for all channel realizations within the uncertainty regions. Consequently, compared to the performance upper bound in ideal cases, i.e., $\tau = 0$, the radar SINR of all schemes has decreased. This is the price to pay to have a robust design and to employ the passive reflecting elements. We emphasize  that the proposed scheme can ensure a sustained sensing quality with increasing $\tau$,  demonstrating its robustness against CSI errors.}	
	\begin{figure}[t]
		\centering
		\includegraphics[width=0.45\textwidth]{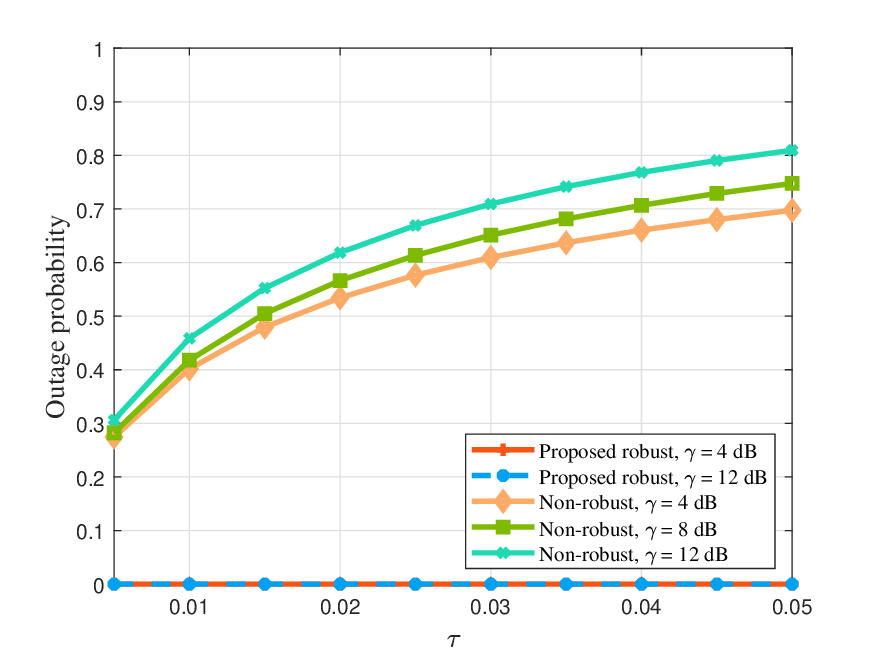}
		\captionsetup{font={normalsize},labelsep=period,singlelinecheck=off}
		\caption{The outage probability versus the communication channel uncertainty level $\tau$ with $\Delta \vartheta= 4^{\circ}$.} 
		\label{outage} 
	\end{figure}%
	
	To further illustrate the robustness of the proposed approach, we plot the outage probability versus the CSI uncertainty level $\tau$ in Fig. \ref{outage}. In particular, the outage probability is defined as the probability that the required QoS $\gamma$ of at least one user is not satisfied. 
	For non-robust designs, we ignore CSI errors and employ the algorithm in~\cite{yang2025joint} to deal with communication SINR constraints. 
	It can be found that, a greater value of QoS requirement $\gamma$ or uncertainty level $\tau$ can lead to a higher outage probability. However, the proposed solution can guarantee that no outage occurs, which demonstrates the robustness of the proposed solution. Here, we would like to emphasize  that even when $\tau$ = 0.005, the outage probability still remains around 30\%. This highlights the critical necessity of developing robust solutions for MA-RIS-DFRC systems.

	\section{Conclusion}
	In this paper, we have proposed a unified robust transceiver design framework of RIS-enhanced DFRC systems with MAs. We considered both imperfect sensing and communication channels. To enhance the sensing performance, we formulated an optimization problem to maximize the minimum radar SINR with regard to all possible locations within the uncertain region subject to the QoS  constraints under the bounded channel error model.
	To deal with the intractable problem, we introduced a two-layer BCD-based algorithm, incorporating FP, SCA, and penalty techniques. We provided a comprehensive analysis of the convergence and computational complexity for the proposed design framework. Simulation results demonstrated that the proposed robust method could significantly improve the  radar performance, and achieve promising balance between the radar and communication performance compared with existing benchmark schemes. 
	
	{There are some promising future research directions. Firstly, in this paper, we modeled both communication and radar channel errors to be bounded, which is relatively conservative. On the other hand, the statistical CSI error model serves as a common alternative for imperfect channel modeling. The proposed optimization algorithm provides a foundational framework for robust designs in MA-enhanced DFRC systems. As such, one of the open problems is to refine and generalize the established optimization framework to tackle statistical CSI errors. Secondly, this work did not account for the positioning inaccuracies of MAs caused by mechanical movement, which can inevitably degrade both sensing and communication performance. Thus, investigating robust transceiver designs that are resilient to imprecise antenna placement is another critical area for future study.}


	\appendices
	\section{Construction of $\nabla\bar{\Gamma}_r(\tilde{\boldsymbol{t}})$ and $\delta_0$ in (\ref{robust3_radar_approximated})}
	Recalling that  $\tilde{\boldsymbol{t}}=[\boldsymbol{t}_1^T,...,\boldsymbol{t}_{N}^T]^T$ and $\boldsymbol{t}_n=[x_n^t,y_n^t]^T$, the gradient vector $\nabla \bar{\Gamma}_r(\tilde{\boldsymbol{t}}) $ and Hessian matrix  $\nabla^2\bar{\Gamma}_r(\tilde{\boldsymbol{t}})$  w.r.t. $\tilde{\boldsymbol{t}}$ are respectively given as follows
		\begin{equation}
			\nabla \bar{\Gamma}_r(\tilde{\boldsymbol{t}}) = \left[ \frac{\partial \bar{\Gamma}_r(\tilde{\boldsymbol{t}})}{\partial x_1^t}, \frac{\partial \bar{\Gamma}_r(\tilde{\boldsymbol{t}})}{\partial y_1^t},..., \frac{\partial \bar{\Gamma}_r(\tilde{\boldsymbol{t}})}{\partial x_{N}^t}, \frac{\partial \bar{\Gamma}_r(\tilde{\boldsymbol{t}})}{\partial y_{N}^t} \right]^T,\label{gradient}
		\end{equation} 	 	
	\begin{equation}
		\nabla^2 \bar{\Gamma}_r(\tilde{\boldsymbol{t}}) =	\begin{bmatrix}
			\begin{matrix}
				\frac{\partial^2 \bar{\Gamma}_r(\tilde{\boldsymbol{t}})}{\partial (x_1^t)^2} & \frac{\partial^2 \bar{\Gamma}_r(\tilde{\boldsymbol{t}})}{\partial x_1^t \partial y_1^t} \\
				\frac{\partial^2 \bar{\Gamma}_r(\tilde{\boldsymbol{t}})}{\partial y_1^t \partial x_1^t} & \frac{\partial^2 \bar{\Gamma}_r(\tilde{\boldsymbol{t}})}{\partial (y_1^t)^2}
			\end{matrix}
			& \cdots & 
			\begin{matrix}
				\frac{\partial^2 \bar{\Gamma}_r(\tilde{\boldsymbol{t}})}{\partial x_1^t \partial y_{N}^t} \\ \frac{\partial^2 \bar{\Gamma}_r(\tilde{\boldsymbol{t}})}{\partial y_1^t \partial y_{N}^t}
			\end{matrix} \\
			\vdots & \ddots & \vdots \\
			\begin{matrix}
				\frac{\partial^2 \bar{\Gamma}_r(\tilde{\boldsymbol{t}})}{\partial y_{N}^t \partial x_1^t} & \frac{\partial^2 \bar{\Gamma}_r(\tilde{\boldsymbol{t}})}{\partial y_{N}^t \partial y_1^t}
			\end{matrix}
			& \cdots &
			\frac{\partial^2 \bar{\Gamma}_r(\tilde{\boldsymbol{t}})}{\partial (y_{N}^t)^2}
		\end{bmatrix}.
		\label{Hessian}
	\end{equation}
	Let us denote the $(i,j)$-th element of $\boldsymbol{D}$ as $d_{i,j} = |d_{i,j}|e^{\jmath\angle d_{i,j}}$, and the $i$-th element of $\boldsymbol{b}_p$ as  $b_{p,i} = |b_{p,i}|e^{\jmath\angle b_{p,i}}$. For notational simplicity, we define that $\kappa_{m,n}^q \triangleq \rho(\boldsymbol{t}_m-\boldsymbol{t}_n,\varphi_q^e,\varphi_q^a)$,  and the calculation on the elements of $\nabla \bar{\Gamma}_r(\tilde{\boldsymbol{t}})$ and $\nabla^2\bar{\Gamma}_r(\tilde{\boldsymbol{t}})$ can be given in (\ref{Obj_cal}), shown at the top of next page, where  
	\begin{figure*}[tp]
		\begin{small}
			\begin{subequations}\label{Obj_cal}
				\begin{alignat}{2}
					&\frac{\partial \bar{\Gamma}_r(\tilde{\boldsymbol{t}})}{\partial x_n^t} = \sum_{p=1}^{P_s} \frac{-4\pi}{\lambda}\sin\varphi_p^e\cos\varphi_p^a|b_{p,n}|\sin\left(\frac{2\pi}{\lambda}\rho(\boldsymbol{t}_n,\varphi_p^e,\varphi_p^a)-\angle b_{p,n}\right)+\sum_{q=1}^{Q}\sum_{m \neq n}^{N}\frac{-4\pi}{\lambda}\sin\varphi_q^e\cos\varphi_q^a c_q|d_{n,m}|\sin\left(\frac{2\pi}{\lambda}\kappa^q_{m,n}+\angle d_{n,m}\right), \\
					&\frac{\partial \bar{\Gamma}_r(\tilde{\boldsymbol{t}})}{\partial y_n^t} = \sum_{p=1}^{P_s} \frac{-4\pi}{\lambda}\cos\varphi_p^e|b_{p,n}|\sin\left(\frac{2\pi}{\lambda}\rho(\boldsymbol{t}_n,\varphi_p^e,\varphi_p^a)-\angle b_{p,n}\right)+\sum_{q=1}^{Q}\sum_{m \neq n}^{N}\frac{-4\pi}{\lambda}\cos\varphi_q^ec_q|d_{n,m}|\sin\left(\frac{2\pi}{\lambda}\kappa^q_{m,n}+\angle d_{n,m}\right),\\
					& \frac{\partial^2 \bar{\Gamma}_r(\tilde{\boldsymbol{t}})}{\partial u_n^t\partial v_n^t} =\sum_{p=1}^{P_s} \frac{-8\pi^2}{\lambda^2}\varphi_p(u_n^t)\varphi_p(v_n^t)|b_{p,n}|\cos\left(\frac{2\pi}{\lambda}\rho(\boldsymbol{t}_n,\varphi_p^e,\varphi_p^a)-\angle b_{p,n}\right)+
					\sum_{q=1}^{Q}\sum_{m \neq n}^{N} \frac{8\pi^2}{\lambda^2}\varphi_q(u_n^t)\varphi_q(v_n^t)c_q|d_{n,m}|\cos\left(\frac{2\pi}{\lambda}\kappa_{m,n}^q+\angle d_{n,m}\right),\\
					& \frac{\partial^2 \bar{\Gamma}_r(\tilde{\boldsymbol{t}})}{\partial u_n^t\partial v_m^t} =  \sum_{q=1}^{Q} \frac{-8\pi^2}{\lambda^2}\varphi_q(u_n^t)\varphi_q(v_m^t)c_q|d_{n,m}|\cos\left(\frac{2\pi}{\lambda}\kappa_{m,n}^q+\angle d_{n,m}\right).
				\end{alignat} 
			\end{subequations} 
		\end{small}
		\hrule
	\end{figure*} $1\leq n\neq m\leq N$, $u_n^t,v_n^t\in\{x_n^t,y_n^t\}, \varphi_p(x_n^t)=\sin\varphi_p^e \cos\varphi_p^a, \varphi_p(y_n^t)=\cos\varphi_p^e,\varphi_q(x_n^t)=\sin\varphi_q^e \cos\varphi_q^a$, and $\varphi_q(y_n^t)=\cos\varphi_q^e$.
	
	 {Next, we consider construction of $\delta_0$. By setting the values of all trigonometric functions to their extrema of $\pm 1$, the theoretical maximum for a single element in the Hessian matrix $\nabla^2 \bar{\Gamma}_r(\tilde{\boldsymbol{t}})$ in~\eqref{Hessian} is found to be
	 	\begin{equation}
	 		\omega =  P_s\frac{8\pi^2}{\lambda^2}b_{\mathrm{max}}+Q(N-1)\frac{8\pi^2}{\lambda^2}c_{\mathrm{max}}d_{\mathrm{max}}.
	 	\end{equation}
	 	Here, $b_{\mathrm{max}}  \triangleq {\mathrm{max}}~ |b_{p,i}|, c_{\mathrm{max}}  \triangleq~ {\mathrm{max}} |c_q|,~ \text{and}~ d_{\mathrm{max}}  \triangleq {\mathrm{max}}~ |d_{i,j}|$, where $1\le i, j \le N, 1\le p \le P_s, 1\le q \le Q$. }
	 	
	 	{Since  $\Vert \nabla^2 \bar{\Gamma}_r(\tilde{\boldsymbol{t}}) \Vert_2\boldsymbol{I}_{2N} \succeq \nabla^2 \bar{\Gamma}_r(\tilde{\boldsymbol{t}})$ and $\Vert \nabla^2 \bar{\Gamma}_r(\tilde{\boldsymbol{t}}) \Vert_2 \leq \Vert \nabla^2 \bar{\Gamma}_r(\tilde{\boldsymbol{t}}) \Vert_F\le 2N\omega$, we can construct $\delta_0$  as 
	 	\begin{equation}
	 		\delta_0 = 2N\omega = \frac{16N\pi^2}{\lambda^2} \left(P_sb_{\mathrm{max}}+Q(N-1)c_{\mathrm{max}}d_{\mathrm{max}}\right).
	 	\end{equation}
	 }

	\section{Construction of $\nabla f_k(\tilde{\boldsymbol{t}})$ and $\delta_k$ in (\ref{upper}) } 
	Let  us denote   the $l$-th element of $\boldsymbol{a}_k$ as $a_{k,l} = |a_{k,l}|e^{\jmath\angle a_{k,l}}$, and define that $\kappa_{k,i,j,l,p}\triangleq-\angle a_{k,l}+\angle a_{k,p} + \frac{2\pi}{\lambda}\rho(\boldsymbol{t}_i,\varphi_l^e,\varphi_l^a)-\frac{2\pi}{\lambda}\rho(\boldsymbol{t}_j,\varphi_p^e,\varphi_p^a) + \angle \boldsymbol{R}_k(i,j)$ and $\mu_{k,i,j,l,p}\triangleq |a_{k,l}||a_{k,p}||\boldsymbol{R}_k(i,j)|$,
	then $f_k(\tilde{\boldsymbol{t}})$ can be reformulated as 
		\begin{align}
			f_k(\tilde{\boldsymbol{t}}) =& \sum_{i=1}^{N}\sum_{l=1}^{L}|a_{k,l}|^2\boldsymbol{R}_k(i,i)
			+\notag\\
			&\sum_{i=1}^{N}\sum_{l=1}^{L-1}\sum_{p=l+1}^{L}2\mu_{k,i,i,l,p}\cos(\kappa_{k,i,i,l,p})+\notag\\
			&\sum_{i=1}^{N-1}\sum_{j=i+1}^{N}\sum_{l=1}^{L}\sum_{p=1}^{L}2\mu_{k,i,j,l,p}\cos(\kappa_{k,i,j,l,p}). 	\label{fkt}	
		\end{align}	
	The derivation of the the gradient vector $\nabla f_k(\tilde{\boldsymbol{t}})$ and Hessian matrix $\nabla^2 f_k(\tilde{\boldsymbol{t}})$ w.r.t. $\tilde{\boldsymbol{t}}$ is similar to~\eqref{gradient} and~\eqref{Hessian}, respectively, and thus omitted for brevity.
	The elements in $\nabla f_k(\tilde{\boldsymbol{t}})$ and $\nabla^2 f_k(\tilde{\boldsymbol{t}})$ can be calculated in (\ref{Com_cal}), 
	\begin{figure*}[tp]
			\begin{subequations}\label{Com_cal}
				\begin{alignat}{2}
					&\frac{\partial f_k(\tilde{\boldsymbol{t}})}{\partial x_i^t} = \sum_{l=1}^{L-1}\sum_{p=l+1}^{L}-\frac{4\pi}{\lambda}\mu_{k,i,i,l,p}\sin(\kappa_{k,i,i,l,p})\left(\sin\varphi_l^e\cos\varphi_l^a-\sin\varphi_p^e\cos\varphi_p^a\right)+ \nonumber\\ 
					&\sum_{j=i+1}^{N}\sum_{l=1}^{L}\sum_{p=1}^{L}-\frac{4\pi}{\lambda}\mu_{k,i,j,l,p}\sin(\kappa_{k,i,j,l,p})\sin\varphi_l^e\cos\varphi_l^a+ \sum_{j=1}^{i-1}\sum_{l=1}^{L}\sum_{p=1}^{L}\frac{4\pi}{\lambda}\mu_{k,j,i,l,p}\sin(\kappa_{k,j,i,l,p})\sin\varphi_p^e\cos\varphi_p^a, \label{gradient_x}\\
					&\frac{\partial f_k(\tilde{\boldsymbol{t}})}{\partial y_i^t} = \sum_{l=1}^{L-1}\sum_{p=l+1}^{L}-\frac{4\pi}{\lambda}\mu_{k,i,i,l,p}\sin(\kappa_{k,i,i,l,p})\left(\cos\varphi_l^e-\cos\varphi_p^e\right)+ \nonumber\\ 
					&\sum_{j=i+1}^{N}\sum_{l=1}^{L}\sum_{p=1}^{L}-\frac{4\pi}{\lambda}\mu_{k,i,j,l,p}\sin(\kappa_{k,i,j,l,p})\cos\varphi_l^e+ \sum_{j=1}^{i-1}\sum_{l=1}^{L}\sum_{p=1}^{L}\frac{4\pi}{\lambda}\mu_{k,j,i,l,p}\sin(\kappa_{k,j,i,l,p})\cos\varphi_p^e,\label{gradient_y} \\
					&\frac{\partial^2 f_k(\tilde{\boldsymbol{t}})}{\partial u_i^t \partial v_i^t}=\sum_{l=1}^{L-1}\sum_{p=l+1}^{L}-\frac{8\pi^2}{\lambda^2}\mu_{k,i,i,l,p}\cos(\kappa_{k,i,i,l,p})\Psi_1(u_i^t)\Psi_1(v_i^t)+\sum_{j=i+1}^{N}\sum_{l=1}^{L}\sum_{p=1}^{L}-\frac{8\pi^2}{\lambda^2}\mu_{k,i,j,l,p}\cos(\kappa_{k,i,j,l,p})\Psi_2(u_i^t)\Psi_2(v_i^t)\nonumber\\
					&+\sum_{j=1}^{i-1}\sum_{l=1}^{L}\sum_{p=1}^{L}\frac{-8\pi^2}{\lambda^2}\mu_{k,j,i,l,p}\cos(\kappa_{k,j,i,l,p})\Psi_3(u_i^t)\Psi_3(v_i^t), \\
					&\frac{\partial^2 f_k(\tilde{\boldsymbol{t}})}{\partial u_i^t \partial v_j^t}=	
					\begin{cases} 
						&\sum_{l=1}^{L}\sum_{p=1}^{L}\frac{8\pi^2}{\lambda^2}\mu_{k,i,j,l,p}\cos(\kappa_{k,i,j,l,p})\Psi_2(u_i^t)\Psi_3(v_j^t),j > i,\\ 
						&\sum_{l=1}^{L}\sum_{p=1}^{L}\frac{8\pi^2}{\lambda^2}\mu_{k,j,i,l,p}\cos(\kappa_{k,j,i,l,p})\Psi_3(u_i^t)\Psi_2(v_j^t),j < i,
					\end{cases} 
				\end{alignat} 
			\end{subequations} 		
		\hrule
	\end{figure*}
	where $u_i^t,v_i^t \in \{x_i^t,y_i^t\},\Psi_1(x_i^t)\triangleq \sin\varphi_l^e\cos\varphi_l^a-\sin\varphi_p^e\cos\varphi_p^a,\Psi_1(y_i^t)\triangleq \cos\varphi_l^e-\cos\varphi_p^e,\Psi_2(x_i^t)\triangleq \sin\varphi_l^e\cos\varphi_l^a,\Psi_2(y_i^t)\triangleq \cos\varphi_l^e,\Psi_3(x_i^t)\triangleq \sin\varphi_p^e\cos\varphi_p^a,$ and $\Psi_3(y_i^t)\triangleq \cos\varphi_p^e$. 
	
	Next, we consider construction of $\delta_k$. Recalling that $\boldsymbol{h}_k = \hat{\boldsymbol{h}}_k +  \Delta\boldsymbol{h}_k$, $\boldsymbol{a}_k$ can be recast as
	\begin{equation}
		\boldsymbol{a}_k = \boldsymbol{\Omega}\hat{\boldsymbol{h}}_k+\boldsymbol{\Omega}\Delta\boldsymbol{h}_k,
	\end{equation}
	where $\boldsymbol{\Omega} = \boldsymbol{\Sigma}^H\boldsymbol{F}(\boldsymbol{s})\boldsymbol{V}^H \in\mathbb{C}^{L\times M}$. Let us denote the $(l,m)$-th entry of $\boldsymbol{\Omega}$ as $\Omega_{l,m} = |\Omega_{l,m}|e^{\jmath\angle\Omega_{l,m}}$, and the $m$-th entry of $\hat{\boldsymbol{h}}_k$ and $\Delta\boldsymbol{h}_k$ as $\hat{h}_{k,m} = |\hat{h}_{k,m}|e^{\jmath\angle\hat{h}_{k,m}}$ and  $\Delta h_{k,m} = |\Delta h_{k,m}|e^{\jmath\angle\Delta h_{k,m}}$, respectively. Then, we have 
		\begin{align}
			|a_{k,l}| &= |\sum_{m=1}^{M}\Omega_{l,m}\hat{h}_{k,m} + \sum_{m=1}^{M}\Omega_{l,m}\Delta h_{k,m}|\notag\\
			&\le  |\sum_{m=1}^{M}\Omega_{l,m}\hat{h}_{k,m}| +| \sum_{m=1}^{M}\Omega_{l,m}\Delta h_{k,m}|\notag\\
			& \le |\sum_{m=1}^{M}\Omega_{l,m}\hat{h}_{k,m}| +\epsilon_k \sum_{m=1}^{M}|\Omega_{l,m}|  \triangleq |\tilde{a}_{k,l}|.
		\end{align}
		{By setting all trigonometric functions to their extreme values, i.e., 1 or -1, the theoretical maximum of a single element in the Hessian matrix  $\nabla^2f_k(\tilde{\boldsymbol{t}})$ is given by 
			\begin{equation}
				\chi_k = \frac{32N\pi^2\upsilon_k}{\lambda^2}\sum_{l=1}^{L}\sum_{p=1}^{L}|\tilde{a}_{k,l}||\tilde{a}_{k,p}|,
			\end{equation}
			where $\upsilon_k \triangleq {\mathrm{max}}~|\boldsymbol{R}_k(i,j)|,1\le i,j\le N$.}
	
	 {Since  $\Vert \nabla^2f_k(\tilde{\boldsymbol{t}}) \Vert_2\boldsymbol{I}_{2N} \succeq \nabla^2f_k(\tilde{\boldsymbol{t}})$ and $\Vert \nabla^2f_k(\tilde{\boldsymbol{t}}) \Vert_2 \leq \Vert \nabla^2f_k(\tilde{\boldsymbol{t}}) \Vert_F\le 2N\chi_k$, we can select $\delta_k$ as 
	 	\begin{equation}
	 		\delta_k = 2N\chi_k =  \frac{64N^2\pi^2\upsilon_k}{\lambda^2} \sum_{l=1}^{L}\sum_{p=1}^{L}|\tilde{a}_{k,l}||\tilde{a}_{k,p}|.
	 \end{equation}}


	\section{Proof of Transformation in (\ref{transformation}) }
	To facilitate the proof, the gradient vectors in (\ref{Com_cal}) first are recast in (\ref{end}),  shown at the top of next page, where
	\begin{figure*}[tp]
			\begin{subequations}\label{end}
				\begin{alignat}{2}
					\frac{\partial f_k(\tilde{\boldsymbol{t}})}{\partial x_i^t} = &   \Re\{\sum_{l=1}^L\sum_{p=1}^La_{k,l}^*\bar{\boldsymbol{R}}_{k,i}(l,p)a_{k,p} + \sum_{j=i+1}^N\sum_{l=1}^L\sum_{p=1}^La_{k,l}^*\bar{\bar{\boldsymbol{R}}}_{k,j,i}(l,p)a_{k,p} + \sum_{j=1}^{i-1}\sum_{l=1}^L\sum_{p=1}^La_{k,l}^*\bar{\bar{\bar{\boldsymbol{R}}}}_{k,j,i}(l,p)a_{k,p}\}
					& = \boldsymbol{a}_k^H\hat{\boldsymbol{A}}_{k,i}\boldsymbol{a}_k,\label{Com_cal_re1}\\
					\frac{\partial f_k(\tilde{\boldsymbol{t}})}{\partial y_i^t} = &\Re\{\sum_{l=1}^L\sum_{p=1}^La_{k,l}^*\bar{\boldsymbol{U}}_{k,i}(l,p)a_{k,p} + \sum_{j=i+1}^N\sum_{l=1}^L\sum_{p=1}^La_{k,l}^*\bar{\bar{\boldsymbol{U}}}_{k,j,i}(l,p)a_{k,p} + \sum_{j=1}^{i-1}\sum_{l=1}^L\sum_{p=1}^La_{k,l}^*\bar{\bar{\bar{\boldsymbol{U}}}}_{k,j,i}(l,p)a_{k,p}\}
					& = \boldsymbol{a}_k^H\hat{\boldsymbol{B}}_{k,i}\boldsymbol{a}_k,\label{Com_cal_re2}
				\end{alignat}
			\end{subequations}
		\hrule
	\end{figure*}
	\begin{small}
		\begin{subequations}
			\begin{alignat}{2}
				\bar{\boldsymbol{R}}_{k,i}(l,p) = &- \frac{2\pi\jmath}{\lambda}\boldsymbol{R}_k(i,i)e^{\frac{\jmath2\pi}{\lambda}(\rho(\boldsymbol{t}_i,\varphi_l^e,\varphi_l^a)-\rho(\boldsymbol{t}_i,\varphi_p^e,\varphi_p^a))}\Psi_1(x_i^t)\notag\\
				\bar{\bar{\boldsymbol{R}}}_{k,j,i}(l,p) = & - \frac{4\pi\jmath}{\lambda}\boldsymbol{R}_k(i,j)e^{\frac{\jmath2\pi}{\lambda}(\rho(\boldsymbol{t}_i,\varphi_l^e,\varphi_l^a)-\rho(\boldsymbol{t}_j,\varphi_p^e,\varphi_p^a))}\Psi_2(x_i^t),\notag\\
				\bar{\bar{\bar{\boldsymbol{R}}}}_{k,j,i}(l,p) =& \frac{4\pi\jmath}{\lambda}\boldsymbol{R}_k(j,i)e^{\frac{\jmath2\pi}{\lambda}(\rho(\boldsymbol{t}_j,\varphi_l^e,\varphi_l^a)-\rho(\boldsymbol{t}_i,\varphi_p^e,\varphi_p^a))}\Psi_3(x_i^t),\notag\\
				\bar{\boldsymbol{U}}_{k,i}(l,p) = &- \frac{2\pi\jmath}{\lambda}\boldsymbol{R}_k(i,i)e^{\frac{\jmath2\pi}{\lambda}(\rho(\boldsymbol{t}_i,\varphi_l^e,\varphi_l^a)-\rho(\boldsymbol{t}_i,\varphi_p^e,\varphi_p^a))}\Psi_1(y_i^t),\notag\\
				\bar{\bar{\boldsymbol{U}}}_{k,j,i}(l,p)= &-\frac{4\pi\jmath}{\lambda}\boldsymbol{R}_k(i,j)e^{\frac{\jmath2\pi}{\lambda}(\rho(\boldsymbol{t}_i,\varphi_l^e,\varphi_l^a)-\rho(\boldsymbol{t}_j,\varphi_p^e,\varphi_p^a))}\Psi_2(y_i^t),\notag\\
				\bar{\bar{\bar{\boldsymbol{U}}}}_{k,j}(l,p,i) = & \frac{4\pi\jmath}{\lambda}
				\boldsymbol{R}_k(j,i)e^{\frac{\jmath2\pi}{\lambda}(\rho(\boldsymbol{t}_j,\varphi_l^e,\varphi_l^a)-\rho(\boldsymbol{t}_i,\varphi_p^e,\varphi_p^a))}\Psi_3(y_i^t).\notag
			\end{alignat} 
		\end{subequations} 
	\end{small}
	Defining that $\hat{\boldsymbol{A}}_{k,i} \triangleq \frac{1}{2}(\bar{\boldsymbol{R}}_{k,i}+\bar{\boldsymbol{R}}_{k,i}^H+\sum_{j=i+1}^N(\bar{\bar{\boldsymbol{R}}}_{k,j,i}+\bar{\bar{\boldsymbol{R}}}_{k,j,i}^H)+\sum_{j=1}^{i-1}(\bar{\bar{\bar{\boldsymbol{R}}}}_{k,j,i}+\bar{\bar{\bar{\boldsymbol{R}}}}_{k,j,i}^H))$ and $\hat{\boldsymbol{B}}_{k,i} \triangleq \frac{1}{2}(\bar{\boldsymbol{U}}_{k,i}+\bar{\boldsymbol{U}}_{k,i}^H+\sum_{j=i+1}^N(\bar{\bar{\boldsymbol{U}}}_{k,j,i}+\bar{\bar{\boldsymbol{U}}}_{k,j,i}^H)+\sum_{j=1}^{i-1}(\bar{\bar{\bar{\boldsymbol{U}}}}_{k,j,i}+\bar{\bar{\bar{\boldsymbol{U}}}}_{k,j,i}^H))$, the gradient vector $\nabla f_k(\tilde{\boldsymbol{t}})$ can be recast   as
	\begin{align}
		\nabla f_k(\tilde{\boldsymbol{t}}) = 	[\boldsymbol{h}_k^H\tilde{\boldsymbol{A}}_{k,1}\boldsymbol{h}_k,\boldsymbol{h}_k^H&\tilde{\boldsymbol{B}}_{k,1}\boldsymbol{h}_k,\dots,\notag\\	&\boldsymbol{h}_k^H\tilde{\boldsymbol{A}}_{k,N}\boldsymbol{h}_k,\boldsymbol{h}_k^H\tilde{\boldsymbol{B}}_{k,N}\boldsymbol{h}_k]^T,
	\end{align}
	where $\tilde{\boldsymbol{A}}_{k,i} = \boldsymbol{V}\boldsymbol{F}(\tilde{\boldsymbol{s}})^H\boldsymbol{\Sigma}\hat{\boldsymbol{A}}_{k,i}\boldsymbol{\Sigma}^H\boldsymbol{F}(\tilde{\boldsymbol{s}})\boldsymbol{V}^H$, $\tilde{\boldsymbol{B}}_{k,i} =\\ \boldsymbol{V}\boldsymbol{F}(\tilde{\boldsymbol{s}})^H\boldsymbol{\Sigma}\hat{\boldsymbol{B}}_{k,i}\boldsymbol{\Sigma}^H\boldsymbol{F}(\tilde{\boldsymbol{s}})\boldsymbol{V}^H,\forall i$. Thus the first-order expansion term in the upper bound of  
	$f_k(\tilde{\boldsymbol{t}})$ can be  recast as
		\begin{align*}
			\nabla f_k(\tilde{\boldsymbol{t}})^T&(\tilde{\boldsymbol{t}}-\tilde{\boldsymbol{t}}^{(l)}) =\notag\\ &\boldsymbol{h}_k^H\underbrace{\{\sum_{i=1}^{N}\tilde{\boldsymbol{A}}_{k,i}(x_i^t-(x_i^t)^{(l)})+\tilde{\boldsymbol{B}}_{k,i}(y_i^t-(y_i^t)^{(l)})\}}_{\triangleq\boldsymbol{\Phi}_k}\boldsymbol{h}_k.
		\end{align*}
	
	With the aid of  $\boldsymbol{h}_k = \hat{\boldsymbol{h}}_k+\Delta \boldsymbol{h}_k$, the proof is completed. \hfill$\Box$

	\balance
	
	\bibliography{reference.bib} 
	\bibliographystyle{IEEEtran} 
	
	%
	
	\vfill
\end{document}